\documentclass[12pt]{article}
\usepackage[right=1.25in,left=1.25in,top=1.1in,bottom=1.1in]{geometry}
\usepackage{mathpazo}
\usepackage{hyperref}
\hypersetup{colorlinks, citecolor=blue, filecolor=blue, linkcolor=blue, urlcolor=blue}
\usepackage{graphicx}
\usepackage{url}
\usepackage[round]{natbib}
\usepackage{amsmath,amsthm} 
\usepackage{engord}
\usepackage{float}
\usepackage{subfig}
\usepackage{pdflscape}
\usepackage[table]{xcolor}
\usepackage{booktabs}
\usepackage{pgfplots}
\usepackage{subcaption}
\usepackage{longtable}
\usepackage{caption}
\pgfplotsset{compat=1.14}
\pgfplotsset{every axis label/.append style={font=\tiny}}
\usepackage[labelsep=period]{caption} %% This switches "Table 1: Title" to "Table 1. Title"

\usepackage{amssymb} %% Necessary, just for the \checkmark command  in tables.
\usepackage{multirow} %% Necessary if we are doing tables in LaTeX

\usepackage{xr}
\usepackage{tcolorbox}

\usepackage{setspace}
\usepackage{subcaption}
\usepackage{sectsty}
\sectionfont{\large}
\subsectionfont{\normalsize}
\subsubsectionfont{\normalsize}

\title{ \vspace*{-2.5cm} \hspace*{-0.5cm}Learning to Regulate: A New Event-Level Dataset of Capital Control Measures
}

\author{Geyue Sun\thanks{Department of Economics, George Washington University.
\href{mailto:geyuesun@gwu.edu}{geyuesun@gwu.edu}} \and Xiao Liu\thanks{Department of Computer Science, University of California, Davis.  \href{mailto: xioliuj@ucdavis.edu}{xioliuj@ucdavis.edu}} \and Tomas Williams\thanks{Department of Economics, George Washington University. \href{mailto: tomaswilliams@gwu.edu}{tomaswilliams@gwu.edu}} \and Roberto Samaniego\thanks{Corresponding Author, Department of Economics, George Washington University.
 \href{mailto:roberto@gwu.edu}{roberto@gwu.edu}}}

\begin{document}

\bgroup
\let\footnoterule\relax

\begin{singlespace}
\maketitle

\begin{abstract}

\noindent We construct a novel event-level Capital Control Measures (CCM) dataset covering 196 countries from 1999 to 2023 by leveraging prompt-based large language models (LLMs). The dataset enables event study analysis and cross-country comparisons based on rich policy attributes, including action type, intensity, direction, implementing entity, and other multidimensional characteristics. Using a two-step prompt framework with GPT-4.1, we extract structured information from the IMF’s Annual Report on Exchange Arrangements and Exchange Restrictions (AREAER), resulting in 5,198 capital control events with 27 annotated fields and corresponding model reasoning. Secondly, to facilitate real-time classification and extension to external sources, we finetune an open-source Meta Llama 3.1-8B model, named \textit{CCM-Llama}, trained on AREAER change logs and final status reports. The model achieves 90.09$\%$ accuracy in category classification and 99.55$\%$ in status prediction. Finally, we also apply the CCM dataset with an empirical application. We conduct an event study on China, Australia and the US. The results show that inward capital control measures significantly reduce fund inflows within one month, and restrictive policies tend to have stronger effects than liberalizing ones, with notable heterogeneity across countries. Our work contributes to the growing literature on the use of LLMs in economics by providing both a novel high-frequency policy dataset and a replicable framework for automated classification of capital control events from diverse and evolving information sources.

\end{abstract}
\end{singlespace}
\thispagestyle{empty}

\clearpage
\egroup
\setcounter{page}{1}

%% Temporary tool to track how this paper is structured. Feel free to comment in or out. 
% \tableofcontents
% \bigskip

%%%%%%%%%%%%%%%%%%%%%%%%%%%%%%%%%%%%%%%%%%%%%%%%%%%%%%%%%%%%%
\section{Introduction\label{sec:introduction}}

\begin{figure}[H]
    \centering
    \includegraphics[width=\textwidth]{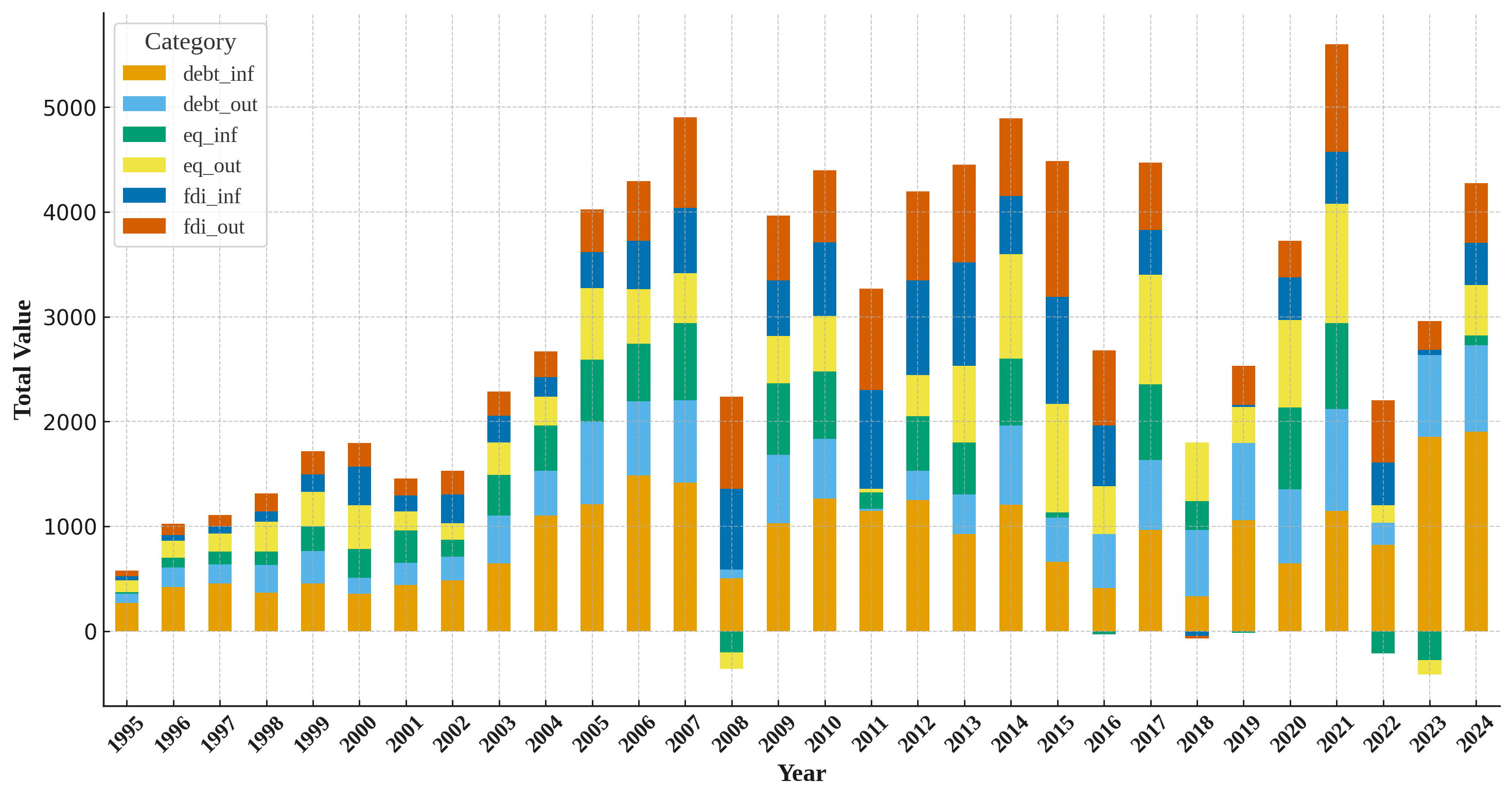}
    \caption{Global Capital Flow Volume}
    \caption*{\small \textit{Notes: Data from the OECD Monthly Capital Flow Dataset \citep{RePEc:oec:dafaaa:2024/02-en}, where $debt~inf$ represents gross portfolio debt inflows, capturing the total value of debt securities acquired by foreign investors; $debt~out$ denotes gross portfolio debt outflows, measuring the volume of domestic investments in foreign debt instruments. Similarly, $eq~inf$ and $eq~out$ correspond to gross portfolio equity inflows and outflows. Additionally, $fdi~inf$ and $fdi~out$ indicate gross foreign direct investment (FDI) inflows and outflows. These variables provide a comprehensive view of international capital mobility across different asset classes.}}
    \label{fig:Global_Capital_Flow}
\end{figure}

\noindent  Capital controls play an essential role in the literature on international capital flows. To mitigate the spillover effects of monetary policy from other countries and reduce financial instability, governments impose restrictions on capital mobility, regulating capital inflows and outflows across different financial accounts. Figure \ref{fig:Global_Capital_Flow} illustrates the absolute volume of various types of capital flows from OECD monthly capital flow dataset. While global capital flows have generally increased since 1995, different financial accounts have exhibited fluctuations: especially during major crises such as the 2008 financial crisis and the COVID-19 pandemic. During those periods, capital control interventions were playing an important role in global capital flows \citep{erten2021capital}. 

There is a large and growing literature examining the effectiveness of capital control policies. \citet{RePEc:nbr:nberwo:18526} finds that, in the face of heightened uncertainty, central banks in developing countries often resort to imposing restrictions on capital outflows. Similarly, \citet{BRUNO2017183} document that capital flow management interventions targeting the banking sector were effective in slowing down banking inflows in a sample of 12 Asia-Pacific economies during the period 2004–2013. More recently, several studies have highlighted the conditional nature of capital control effectiveness. \citet{NBERw26558} provide a comprehensive survey and argue that the success of capital controls depends critically on factors such as institutional quality, the timing of intervention, and the type of capital flow targeted. Similarly, \citet{bergant2024dampening} show that, relative to macroprudential regulation, capital controls have a weaker impact on macroeconomic stability. Focusing specifically on capital outflows, \citet{NBERw32877} investigate capital controls on outflows (CCOs) during periods of financial distress and find that government credibility and reputation are essential for their successful implementation. And \citet{FORBES2015S76} also find that while certain capital flow management measures—such as removing outflow controls—may help mitigate real exchange rate appreciation, most CFMs have limited impact on key macro-financial variables and often fall short of achieving their stated policy objectives. In addition, recent work also explores the interaction between capital controls and other policy domains. \citet{LLOYD2024103965}, for instance, show that capital control measures and trade policies are closely linked: optimal capital controls tend to be smaller when trade policy is constrained by agreements such as FTAs, though under certain conditions they may be larger depending on macroeconomic fundamentals. 

Thus, the existing literature underscores the importance of heterogeneity in capital control measures in shaping cross-border capital flows. By looking into the data used in the empirical analysis, most empirical studies rely on index-based approaches to capture the presence or intensity of capital controls. A prominent example is the dataset constructed by \citet{d814d833-3bfa-385a-90ba-0331a8fc90bb}, \citet{e44f2e54-ab1a-30a8-8fa3-c3bd8f4cef47}, who use the IMF’s Annual Report on Exchange Arrangements and Exchange Restrictions (AREAER) to produce binary indicators of capital control measures by asset class and country. These indicators are derived from the “status” column in the AREAER reports, which flags the presence of capital restrictions using a “Yes/No” format.\footnote{The AREAER includes a column labeled "status" that provides a binary indicator of whether a given restriction is in place.} Based on this information, they construct annual indices taking values between 0 and 1 to reflect the existence of restrictions across categories. The binary construction implicitly treats all capital control measures as equally binding or consequential, masking the variation in intensity, scope, and implementation detail. Related indices such as the Chinn-Ito index \citep{NBERw11370} and its extensions \citep{RePEc:pal:imfstp:v:56:y:2009:i:1:p:222-238} also rely on binary information from the AREAER to produce aggregate measures of financial openness. While useful for identifying broad trends, these indices abstract from the underlying heterogeneity of capital control instruments and do not distinguish between liberalizing and restrictive interventions or account for conditional policies.

As discussed above, current capital control indices face two key limitations. First, they are typically available only at low frequency; second, they provide limited information about the characteristics of the underlying policy interventions. By relying on binary classification, existing indices may obscure meaningful variation in the content of capital control measures, treating qualitatively distinct interventions as equivalent. While this approach is tractable, it risks overlooking important differences in the design, scope, and enforcement of capital controls. To contribute to a more detailed understanding of policy dynamics, we construct CCM Dataset, a new event-level dataset enriched with multiple policy characteristics by leveraging textual analysis using prompt-based large language models (GPT-4.1).\footnote{Our dataset relies on the textual description of each policy to provide attributes such as whether a measure is restrictive, conditional, or liberalizing; whether it is implemented at the national or sub-national level; and which sectors or countries are targeted, among other features.} By integrating these dimensions, CCM dataset captures event-level capital intervention measures and characterizes each policy action, offering a more granular and multidimensional view of capital control measures over time.

Another major challenge in current capital control research is the costly and labor-intensive nature of data construction. Traditional indices rely on manual review of the AREAER reports, where researchers must examine each policy announcement to determine whether it belongs to a particular category, constitutes an exemption, or reflects potential misclassifications. This procedure is both time-consuming and expertise-intensive, and is tied to annual reporting cycles, which delay access to up-to-date information for relevant organizations and researchers and restrict the frequency of available data. Furthermore, each regulatory action must be individually assessed and coded, adding considerable complexity to the process. To address these limitations, leveraging generative AI, we finetune an open-source LLM called \textit{CCM-Llama}, which offers a novel yet practical alternative to the previous labor-intensive data construction by automating the assignment of capital intervention policies to the appropriate IMF-aligned categories. This approach improves both the efficiency and scalability of capital control measures data collection, while enabling more detailed and real-time and multi-source policy analysis.

The remainder of the paper is organized as follows. Section~\ref{sec:promptLLM} outlines the textual analysis methodology using prompt-based LLMs and presents key stylized facts derived from the CCM dataset. Section~\ref{sec:finetuning} details the finetuning process of an open-source LLM and evaluates its performance in classifying capital control measures. Section~\ref{sec:empirical} illustrates an event-study application of the CCM dataset, examining the effectiveness of capital control policies on global fund flows. Section~\ref{sec:conclusion} concludes.

% In addition, existing datasets often lack detailed, event-level documentation, as AREAER typically records only a limited number of changes each year. By integrating the capital intervention dataset from Global Trade Alert, we enhance the dataset with more granular, high-frequency information, capturing short-term policy shifts and offering a more accurate picture of capital control dynamics over time.

%%%%%%%%%%%%%%%%%%%%%%%%%%%%%%%%%%%%%%%%%%%%%%%%%%%%%%%%%%%%%%%%%%%

\newpage
\section{Textual Analysis of Capital Control Measures with Prompt-Based LLMs} \label{sec:promptLLM}

In this section, we leverage textual data on capital controls from the IMF's \textit{Annual Report on Exchange Arrangements and Exchange Restrictions} (AREAER) and apply pre-trained large language models (LLMs) to extract multidimensional information about capital interventions. The AREAER provides narrative descriptions of policy changes for each country, including the nature, scope, and context of capital account regulations. By combining domain-specific prompts with the reasoning capabilities of LLMs, we instruct the LLM to identify key features of each intervention. This approach enables us to construct a structured, event-level dataset that captures both the timing and characteristics of capital control actions at a daily frequency.

\subsection{Textual Data Description}
\label{sec:areaer_des}
Our main source of country-level textual data comes from the IMF’s \textit{Annual Report on Exchange Arrangements and Exchange Restrictions} (AREAER), spanning the years 1999 to 2023 for 196 countries. The AREAER provides standardized, annually updated documentation across 12 main sections, covering topics such as national exchange rate regimes and detailed legal and administrative descriptions of capital account regulations. Each report contains over 250 subcategories per year, offering granular policy information for a wide range of instruments and regulatory dimensions.

In addition to these structured elements, the AREAER includes narrative descriptions of policy changes that document the adoption, amendment, or removal of specific measures \footnote{Refer to the AREAER online platform:\href{https://www.elibrary-areaer.imf.org/Pages/AboutUs.aspx}{About Us}}. These change records are organized by section and typically report the effective date of implementation. Precise timing is essential for conducting event-study analyses, as it allows researchers to align policy interventions with market responses or macroeconomic outcomes at high temporal resolution. The combination of consistency in reporting and the richness of unstructured description text makes the AREAER particularly well suited for information extraction and classification using large language models \footnote{Between 1999 and 2022, the structure of the AREAER underwent several revisions, including the discontinuation of certain categories and the introduction of new ones. We account for these structural changes by mapping discontinued categories to consistent equivalents where possible to maintain comparability.}.

\begin{figure}[H]
    \centering
    \includegraphics[width=\textwidth]{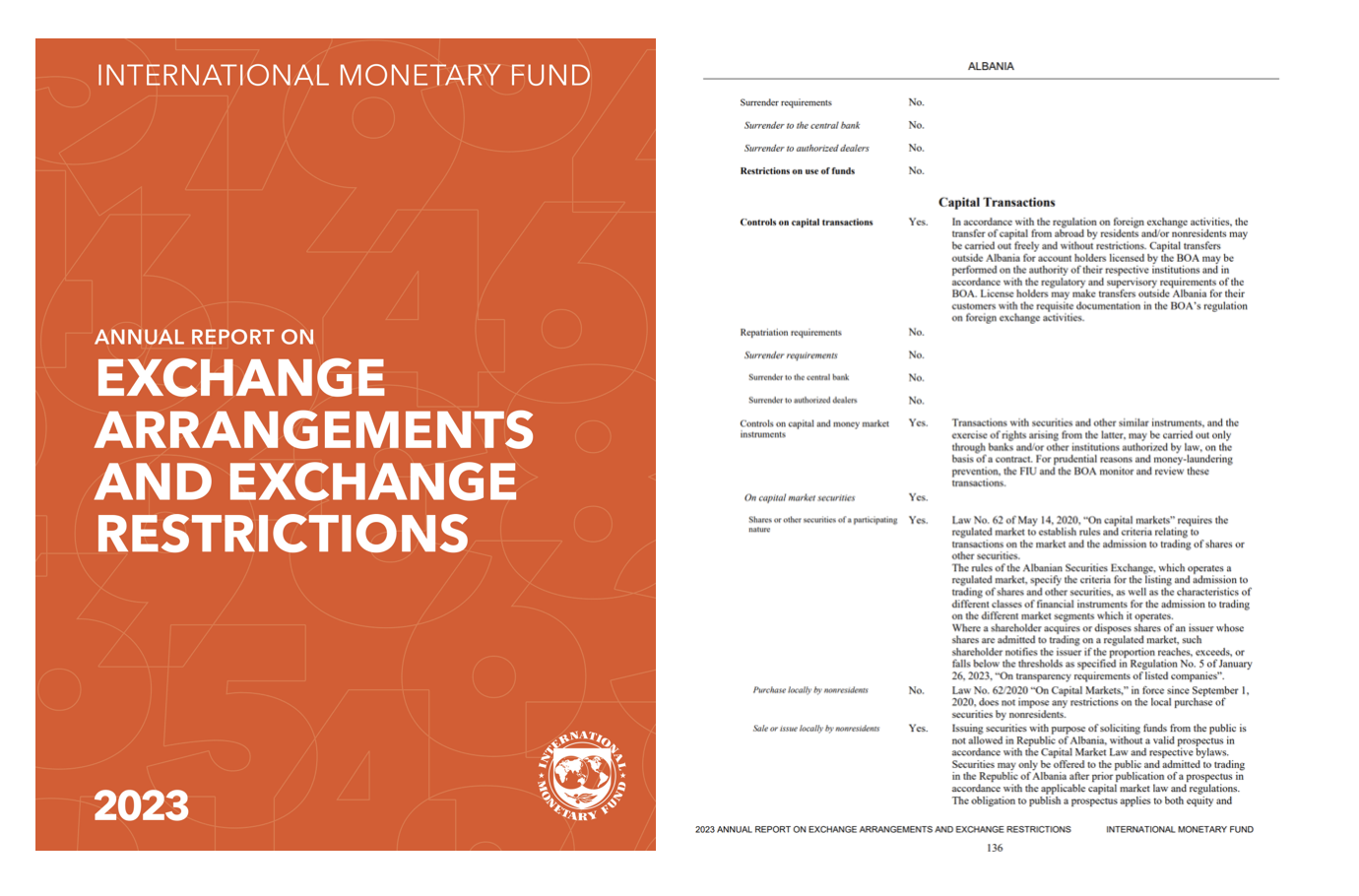}
    \caption{Example of AREAER Reports}
    \caption*{\small \textit{Notes: This picture is an example of 2023 AREAER report from IMF online platform: \hyperlink{https://www.elibrary-areaer.imf.org/Documents/YearlyReport/AREAER_2023.pdf}{click here}}}
    \label{fig:AREAER_report}
\end{figure}

\begin{table}[ht]
\centering
\footnotesize
\caption{Total Count and Word Count of Capital Control Descriptions by Year}
\label{tab:capital_control_distribution}
\begin{tabular}{r|rr|rr}
\toprule
\textbf{Year} & \multicolumn{2}{c|}{\textbf{Panel A: Description on Final Report }} & \multicolumn{2}{c}{\textbf{Panel B: Description on Yealy Changes}} \\
             & Count & Word Count & Count & Word Count \\
\midrule
1999 & 1111 & 9872 & --- & --- \\
2000 & 1116 & 8128 & 59 & 1615 \\
2001 & 1116 & 8531 & 75 & 1991 \\
2002 & 1122 & 8867 & 78 & 2327 \\
2003 & 1122 & 8970 & 90 & 2424 \\
2004 & 1122 & 9680 & 79 & 1915 \\
2005 & 1122 & 9962 & 30 & 908 \\
2006 & 1128 & 11505 & 102 & 3079 \\
2007 & 1128 & 12083 & 145 & 4671 \\
2008 & 1128 & 11827 & 121 & 4082 \\
2009 & 1134 & 13639 & 92 & 3708 \\
2010 & 1140 & 15628 & 85 & 4059 \\
2011 & 1140 & 17022 & 94 & 6369 \\
2012 & 1146 & 17359 & 97 & 4511 \\
2013 & 1140 & 16149 & 143 & 7938 \\
2014 & 1146 & 18512 & 162 & 7545 \\
2015 & 1152 & 22779 & 222 & 13852 \\
2016 & 1146 & 19444 & 270 & 19894 \\
2017 & 1152 & 20447 & 337 & 21686 \\
2018 & 1152 & 20097 & 241 & 14701 \\
2019 & 1152 & 19486 & 414 & 28793 \\
2020 & 1158 & 20565 & 347 & 21825 \\
2021 & 1164 & 21441 & 385 & 25115 \\
2022 & 1164 & 21697 & 436 & 21306 \\
2023 & ---  & ---   & 343 & 22218 \\
\bottomrule
\end{tabular}
% \begin{tablenotes}
% \footnotesize
% \item \textit{Notes:} Panel A reports the total number of AREAER capital control descriptions by year. Panel B is the changes policies 
%  in each years. We also show the data geography distribution, and also the subcategory distribution.   
% \end{tablenotes}
\end{table}

As demonstrated in Figure~\ref{fig:AREAER_report}, the AREAER is updated annually on the IMF’s official website\footnote{The most recent AREAER report covers all policy changes through 2022 and was published in 2023.}. Each report includes narrative descriptions of capital transaction controls and financial sector measures implemented by individual countries. These entries are organized by category index, category name, a corresponding textual description, and a binary status indicator. Table~\ref{tab:capital_control_distribution} summarizes the distribution of capital control measures and recorded changes across the full set of AREAER reports used in our analysis.

\noindent\paragraph{Category} We follow the IMF AREAER classification under Section XI.A, which outlines controls on cross-border capital transactions. Our focus is on subcategories XI.A.2 through XI.A.7, covering various instruments and transaction types. Specifically, XI.A.2 addresses controls on capital and money market instruments such as equities, bonds, and mutual funds; XI.A.3 covers derivatives and other financial instruments; XI.A.4 pertains to credit operations, including financial and commercial credits as well as guarantees; XI.A.5 focuses on direct investment, including both inward and outward FDI and the liquidation of such investments; and XI.A.7 relates to real estate transactions. These subcategories are further disaggregated by investor residency (resident vs. nonresident), direction of capital flow (inward vs. outward), and the nature of the instrument.  The complete list of 45 category indices and corresponding names is provided below.\footnote{The indices and names follow the official classification used in the IMF AREAER.}

\begin{tcolorbox}[colback=gray!10, colframe=white!50]
\footnotesize
\begin{verbatim}

 
("XI.A", "Controls on capital transactions"),  
("XI.A.2", "Controls on capital and money market instruments"),
("XI.A.2.a", "On capital market securities"),
("XI.A.2.a.1", "Shares or other securities of a participating nature [eq]"),
("XI.A.2.a.1.i", "Purchase locally by nonresidents [eq_plbn]"),
("XI.A.2.a.1.iv", "Sale or issue abroad by residents [eq_siar]"),
("XI.A.2.a.1.iii", "Purchase abroad by residents [eq_pabr]"),
("XI.A.2.a.1.ii", "Sale or issue locally by nonresidents [eq_siln]"),
("XI.A.2.a.2", "Bonds or other debt securities [bo]"),
("XI.A.2.a.2.i", "Purchase locally by nonresidents [bo_plbn]"),
("XI.A.2.a.2.iv", "Sale or issue abroad by residents [bo_siar]"),
("XI.A.2.a.2.iii", "Purchase abroad by residents [bo_pabr]"),
("XI.A.2.a.2.ii", "Sale or issue locally by nonresidents [bo_siln]"),
("XI.A.2.b", "On money market instruments [mm]"),
("XI.A.2.b.1", "Purchase locally by nonresidents [mm_plbn]"),
("XI.A.2.b.4", "Sale or issue abroad by residents [mm_siar]"),
("XI.A.2.b.3", "Purchase abroad by residents [mm_pabr]"),
("XI.A.2.b.2", "Sale or issue locally by nonresidents [mm_siln]"), 
("XI.A.2.c", "On collective investment securities [ci]"),
("XI.A.2.c.3", "By residents to nonresidents [cio]"),
("XI.A.2.c.1", "By nonresidents to residents [cii]"),
("XI.A.3", "Controls on derivatives and other instruments"),
("XI.A.3.a", "Purchase locally by nonresidents"),
("XI.A.3.b", "Sale or issue locally by nonresidents"),
("XI.A.3.c", "Purchase abroad by residents"),
("XI.A.3.d", "Sale or issue abroad by residents"), 
("XI.A.4", "Controls on credit operations"),
("XI.A.4.b", "Financial credits [fc]"),
("XI.A.4.b.1", "By residents to nonresidents [fco]"),
("XI.A.4.b.2", "By nonresidents to residents [fci]"), 
("XI.A.4.a", "Commercial credits"),
("XI.A.4.a.1", "By residents to nonresidents"),
("XI.A.4.a.2", "To residents from nonresidents"),
("XI.A.4.c", "Guarantees, sureties, and financial backup facilities"),
("XI.A.4.c.1", "By residents to nonresidents"),
("XI.A.4.c.2", "To residents from nonresidents"),
("XI.A.7", "Controls on real estate transactions"),
("XI.A.7.a", "Purchase abroad by residents"),
("XI.A.7.b", "Purchase locally by nonresidents"),
("XI.A.7.c", "Sale locally by nonresidents"),
("XI.A.5", "Controls on direct investment [di]"),
("XI.A.5.a", "Outward investment [dio]"),
("XI.A.5.b", "Inward direct investment [dii]"),
("XI.A.5.c", "Liquidation of direct investment [ldi]")
    
\end{verbatim}

\end{tcolorbox}

\noindent\paragraph{Description} Under each category, the AREAER provides a narrative description of relevant policy announcements that became effective on specific dates within each country. As shown in Table~\ref{tab:example_policy_entries}, by reading the textual reports, we find the description of capital control policies documented as the follows. These entries clearly document changes in regulations or legal provisions, often citing the official titles of laws or decrees issued by central banks or foreign exchange authorities.

We use this rich textual information as the basis for extracting detailed capital control interventions. For each event, we identify key attributes such as the type of action taken, its intensity, policy direction (e.g. liberalizing or tightening), administrative level (e.g., national or sub-national), and whether the measure is related to trade or national security sanctions.\footnote{We provide additional details on the prompt-based methodology and the structure of the resulting dataset in the following Section~\ref{sec:methodology}.}

\begin{table}[ht]
\centering
\footnotesize
\caption{Examples of Capital Control Policy Descriptions with Status}
\label{tab:example_policy_entries}
\begin{tabular}{llp{4cm}p{4.8cm}l}
\toprule
\textbf{Date} & \textbf{Country} & \textbf{Category} & \textbf{Description} & \textbf{Status} \\
\midrule
01/01/2014 & Germany & Provisions specific to commercial banks and other credit institutions & Regulation (EU) No. 575/2013 on Prudential Requirements for Credit Institutions and Investment Firms introduced measures that may affect cross-border operations and capital adequacy within the EU banking framework. & yes \\
02/07/2013 & China & Repatriation requirements & Proceeds from the issuance of shares by an overseas-listed Chinese company must be repatriated within a prescribed time and used in accordance with approved purposes. & yes \\
09/14/2016 & United States & Inward direct investment & The sanction program that restricted certain investments related to North Korea remained in force, affecting financial transactions involving designated entities. & yes \\
\bottomrule
\end{tabular}
\end{table}

\noindent\paragraph{Status} The final yearly textual report includes a “status” field, which provides a binary response as "Yes/No", indicating whether the policy described in the text qualifies as a capital control and alters the country’s regulatory stance in the corresponding category, as shown in Table \ref{tab:example_policy_entries}. For example, if a liberalizing measure eliminates all restrictions in a specific category, the status is recorded as 'No'. Conversely, if a new policy reinforces existing restrictions without changing the general regime, the status remains 'Yes', consistent with the previous year. Since the content reports from AREAER do not include the exact implementation dates of policy changes, we infer the corresponding status of each measure by matching the reported changes in a given year with the final status reported in the following year’s content report, the detail data processing could be refer to Figure \ref{fig:data_collect} in the finetuning section.

This binary indicator plays a central role in conventional capital control indices, which typically use the final status to quantify a country’s level of capital account restrictiveness. However, this approach overlooks important dynamics—such as the intensification of controls within a category—that do not alter the binary status but nevertheless reflect meaningful policy shifts. This is a key contribution of our paper: by providing an event-level dataset, we expose the granularity that conventional indices miss. In future research, this dataset could serve as the foundation for constructing more dynamic and precise capital control indices that account for both the frequency and depth of policy interventions.

\subsection{Methodology} \label{sec:methodology}
We begin by using frontier pre-trained large language models (LLMs) to extract multidimensional characteristics from the rich textual descriptions of capital control policies. Specifically, we utilize the API-based GPT-4.1 \citep{gpt412025} model, given its strong performance in processing regulatory text, broad knowledge base, and advanced reasoning capabilities. The model is used for initial information extraction via OpenAI’s API. In contrast, later Section \ref{sec:finetuning} leverages finetuned, locally hosted LLMs (e.g., Meta Llama 3.1 \citep{grattafiori2024llama}) to enable more customized classification and model adaptation.

Our methodology is based on large-scale inference from unstructured policy text using GPT-4.1 \citep{gpt412025}, a proprietary model with state-of-the-art comprehension. To ensure robustness, we set the temperature level equal to 0, and cross-validate selected outputs with alternative models, including GPT-3.5, and confirm that results are qualitatively consistent across versions.

The extraction process is structured using a two-part prompt design. First, a system prompt defines the classification schema and provides detailed instructions for mapping policy content to structured attributes (e.g., implementation level, firm targeting, policy type). Second, the unstructured textual description is submitted as the user prompt. This structure ensures consistency across documents and facilitates scalable, standardized output. Model inference is executed via OpenAI’s API using multi-thread processing. To enhance reproducibility and eliminate output randomness. Extracted results are parsed and stored in structured formats (e.g., JSON or CSV) for downstream empirical analysis. All API-based calls are conducted in a secure cloud environment, with rate-limited batching to comply with OpenAI usage constraints and cost-efficiency considerations.

In the first step, we prompt the GPT-4.1 \citep{gpt412025} model with background knowledge about the structure of the AREAER annual reports. To accomplish this, we provide the model with sample reports and instruct it to read and interpret the layout and content of each section. This step ensures that the model understands the institutional and legal context in which capital control measures are described.

At the same time, we incorporate definitions and classification standards from two complementary data sources. First, we reference the Dataset of Capital Flow Management Measures (CFMs), which defines CFMs as policy tools aimed at influencing cross-border capital movements, including both inflows and outflows \citep{NBERw32877}. Second, we draw on the structure of the Global Trade Alert (GTA) dataset, which classifies state interventions using a three-tier system: the country act grouping, the state act, and the intervention. In our implementation, we focus on the country act and intervention levels, consistent with how GTA organizes and collects policy information \citep{RePEc:cup:wotrrv:v:8:y:2009:i:04:p:607-609_99}.

\begin{tcolorbox}[colback=gray!10, colframe=white!50]
\footnotesize
\begin{verbatim}
system_prompt = """
You are a senior policy analyst. Your job is to extract structured information 
from IMF Exchange Arrangements and Exchange Restrictions (AREAER) reports. 
The example of reports is as follows. And the textual report includes the 
capital control policies in countries with detailedd descriptions: 

{"index": "X.D.2.", "category": "Inward direct investment", "code": "172", 
"status": "yes", "description": "Investments require prior approval and 
are administered by the Investment Committee. The law stipulates that 
foreign investment in the Islamic State of Afghanistan can take place only
through joint ventures, with foreign participation not exceeding 49%, and 
that an investment approved by the Investment Committee requires no further
license in order to operate in the Islamic State of Afghanistan. The Foreign
and Domestic Private Investment Law includes the following provisions: 
(1) income tax exemption for four years (six years outside Kabul province),
beginning with the date of the first sale of products resulting from the new
investment; (2) exemption from import duties on essential imports (mainly for
capital goods)}
"""
\end{verbatim}

\end{tcolorbox}

Once the model is able to understand the structure of the AREAER reports and recognize capital control–related categories and policy descriptions, we proceed to prompt the LLM to read individual entries and extract the following characteristics from the text. To facilitate manual verification of the assigned categories, we also instruct the model to generate a brief justification for each classification and to cite the corresponding sentence(s) from the original description. This step helps ensure the transparency and interpretability of the model's outputs, and allows researchers to audit the classification process when necessary.

\begin{tcolorbox}[colback=gray!10, colframe=white!50]
\footnotesize
\begin{verbatim}
You are a senior policy analyst. Your job is to extract structured information
from IMF capital control descriptions.
Return all results as a JSON object with specific fields. Use precise and short
terms.
If a field is not explicitly mentioned, return null or false.
For each extracted field, also include a new field called '[Field]_source_
sentence' showing the exact sentence from the original description that
supportsyour extraction.

For 'Action', classify by using the verb words for applying the policies:

- "Action": Classify the policy's primary regulatory move into one of the 
following **10 fixed categories**.
Choose the most important action described in the text. If multiple actions
are present, select the one with the greatest policy impact. Return only
one of the following values:
    - prohibit: Explicitly bans a type of transaction or capital flow.
      Example: "Residents are prohibited from acquiring foreign bonds."
    - limit: Imposes a cap or threshold on volume, frequency, or eligible 
    entities.
      Example: "A ceiling was placed on foreign portfolio investment."
    - suspend: Temporarily halts or freezes an activity that was previously
    allowed.
      Example: "Licensing of outward investments was suspended."
    - require approval: Allows the activity only if prior approval or 
    registration is granted.
      Example: "All real estate purchases by nonresidents require prior 
      approval."
    - subject to quota: Allows the activity but restricts it under a quota 
    or allocation.
      Example: "Foreign exchange purchases for education are subject to quota."
    - permit: Explicitly authorizes or legalizes an activity that was 
    previously restricted.
      Example: "Corporates are permitted to invest abroad in joint ventures."
    - remove: Cancels a previous restriction or approval requirement.
      Example: "The requirement for prior approval was removed."
    - ease: Makes an existing restriction more flexible (e.g., expanding 
    coverage, simplifying approval).
      Example: "Approval procedures were eased for investment abroad."
    - amend: Modifies the wording or scope of existing regulation without 
    clearly changing its restrictiveness.
      Example: "The regulation was amended to update procedural definitions."
    - clarify: Provides clarification or interpretation of existing rules 
    without changing legal effect.
      Example: "The scope of 'resident' was clarified to include offshore 
      subsidiaries."
...
\end{verbatim}

\end{tcolorbox}

\begin{tcolorbox}[colback=gray!10, colframe=white!50]
\footnotesize
\begin{verbatim}
For 'Action Intensity', classify only into the following categories based on 
meaning:
- restrictive: Used when the policy limits or prohibits cross-border 
transactions.Common cues: prohibit, ban, restrict, suspend, limit, impose, 
freeze.
  Example: "Residents are no longer allowed to purchase foreign securities."
- liberalizing: Used when the policy removes, relaxes, or eases controls. Cues: 
allow, remove, lift, permit, ease, liberalize.
  Example: "Restrictions on nonresidents purchasing local bonds were lifted."
- conditional: Used when the policy allows actions only under specific 
conditions (approval, quota, limits).
  Example: "Foreign investment is allowed, subject to approval by the 
  Ministry."
- neutral: Used when the change is administrative or direction is unclear.
  Example: "The central bank clarified reporting procedures for offshore 
  transactions."
  
For 'Action Direction', classify only as:
- inward: Affects inflows into the country.
- outward: Affects outflows from the country.
- both: Simultaneously affects inward and outward flows.
- undefined: Direction not specified or unclear.
For 'Action Level', classify only as:
- supranational: Decision made by a regional or international body (e.g., EU).
- national: Implemented by central government.
- subnational: Implemented by local or provincial authority.
- undefined: Level of authority not specified.
For 'Instrument', classify by using the policies tools (e.g., foreign exchange, 
bank credit,bank reserve requirement)
For 'Actor', classify by using the policies makers (e.g., central bank, 
government, commercial banks)
For 'Beneficiary', classify by using the policies aiming groups of investor 
(e.g., exporters, travelers)
For 'Condition', classify by additional requirement (e.g., without approval,
only on request,subject to approval)
- without  - with  (e.g., only on request,subject to approval)
For 'Target Country', identify the country who been targeted with the policies
For 'Target Industry', identify the industry which been targeted with the 
policies
For 'Is Trade Policy', return true if the measure affects trade in goods or 
tariffs (e.g., includes words like import, export, quota, tariff).
For 'Is Sanction', return true if the measure involves bans, freezes, or 
disallowing actions typically against countries or individuals.
For 'Is National Security', return true if the justification or mechanism 
refers to security, terrorism, or national interest.
"""
\end{verbatim}

\end{tcolorbox}

We initially store the model outputs in JSON format and subsequently convert them into CSV files upon completion of the LLM processing. After extracting relevant information from each policy description in the AREAER, we match the resulting records with metadata including the policy date, country name, status, and categorical identifiers such as geographic region, income level, and sub-income classification.

\subsection{Capital Control Measures Dataset: Stylized Facts }
The structure of the Capital Control Measures (CCM) dataset is summarized in Table~\ref{tab:capital_control_intervention}. Each observation represents a capital control policy intervention at the daily level, classified across a rich set of dimensions. In addition to basic identifiers such as the date of implementation, reporting country, and full policy description, each entry is aligned with the hierarchical IMF AREAER classification system through a category index and category name.

Our CCM dataset captures a wide array of characteristics for each policy event, including the type of action (e.g., prohibit, permit), its intensity (restrictive, conditional, liberalizing), direction (inward, outward, or both), and jurisdictional level (national, subnational, or supranational). It further records contextual details such as the instrument used, the implementing authority, affected sectors or entities, and whether the policy relates to trade, sanctions, or national security. To ensure interpretability, each entry also includes the original IMF AREAER status (Yes/No) and a justification generated by a large language model, citing relevant text and reasoning.

\begin{table}[htbp]
\footnotesize
\centering
\caption{Structure of the Capital Control Measures Dataset}
\label{tab:capital_control_intervention}
\scalebox{0.95}{
\begin{tabular}{ll}
\toprule
\textbf{Field} & \textbf{Description} \\
\midrule
Year & The implemented year of the capital control policies \\
IFS code & Country code \\
Country & Name of the reporting country \\
Region & Region of the reporting country \\
Income group & Income group of the reporting country based on World Bank \\
Income subgroup & Income subgroup of the reporting country based on World Bank \\
Index Code & Index code provided by IMF \\
Category Index & Hierarchical index consistent with IMF AREAER classification \\
Category & Name of the capital control category (e.g., XI.A.2.a.1.i) \\
Date & Effective date of the intervention \\
Description & Full text description of the capital control measure \\
Date of Retroactive Changes  &  The date if there is a date for canceling this policy  (if applicable) \\
Action & Verbs describing the measure (e.g., prohibit, permit, require approval) \\
Action Intensity & Restrictive / Conditional / Liberalizing / Neutral \\
Action Direction & Inward / Outward / Both / Undefined \\
Action Level & Supranational / National / Subnational / Undefined \\
Instrument & Policy tool involved (e.g., foreign exchange, credit, reserve requirement) \\
Actor & Entity implementing the policy (e.g., central bank, government) \\
Condition & With (e.g., subject to approval) / Without (e.g., no approval required) \\
Beneficiary & Group intended to benefit or be affected (e.g., exporters, travelers) \\
Target Country & Country explicitly targeted by the measure (if applicable) \\
Target Industry & Sector targeted by the policy (if applicable) \\
Target Industry & Sector targeted by the policy (if applicable) \\
Limit/Threshold & Limits number or threshold of the policy  (if applicable) \\
Is Trade Policy & True if measure relates to trade in goods or tariffs \\
Is Sanction & True if measure involves bans, freezes, or restricted parties \\
Is National Security & True if measure references security or national interest \\
% Status  & Yes/No indicating IMF AREAER status \\
LLM Reasoning & Cited phrase and rationale reasoning extracted by LLMs \\
\bottomrule
\end{tabular}

}

\end{table}

\begin{figure}[H]
    \centering
    \includegraphics[width=\textwidth]{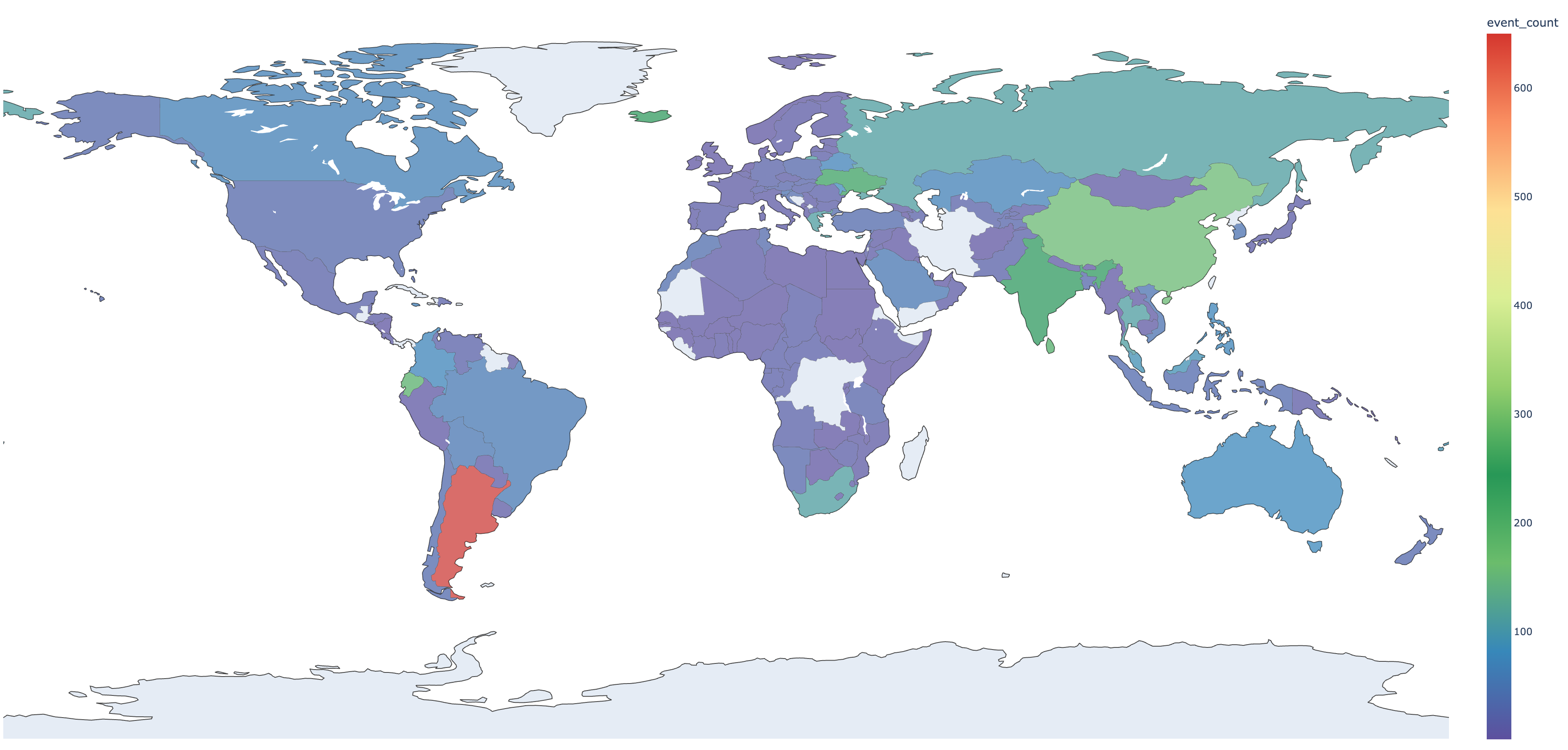}
    \caption{Total Capital Control Measures Distribution Map}
    \label{fig:Distribution_map}
    \end{figure}

In summary, the CCM dataset covers 45 types of categorized capital control interventions across 196 countries from 1999 to 2023.\footnote{In the finetuning section (Section \ref{sec:finetuning}), the training dataset covers data only up to 2022, as the final AREAER report for each year is required to determine the status conditions. The AREAER report for 2023 has not yet been released as of the time of writing.} It provides event-level information at a daily frequency on policy actions implemented globally to regulate cross-border capital flows. Figure~\ref{fig:Distribution_map} presents the geographic distribution of capital control measures across different world regions. The dataset captures policy interventions from nearly all countries covered by the IMF AREAER, offering broad global coverage. In total, the dataset includes 5,198 documented intervention events.

\begin{figure}[H]
    \centering
    \includegraphics[width=0.9\textwidth]{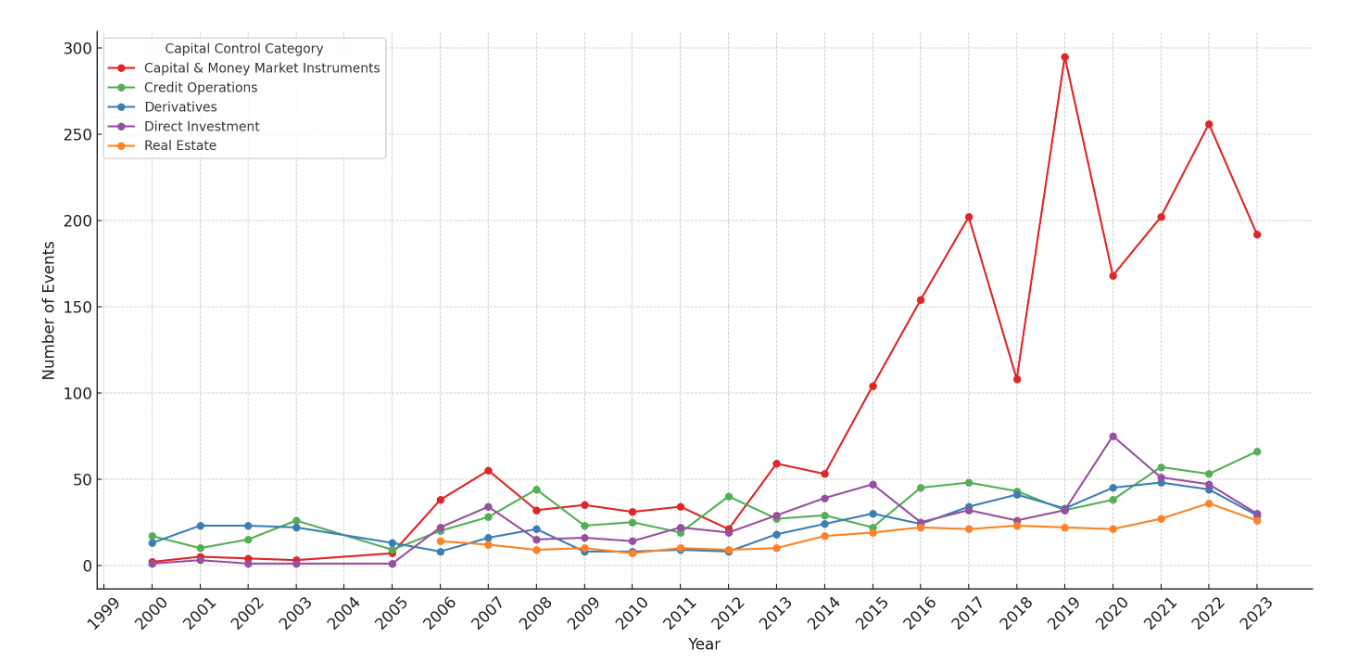}
    \caption{Total Capital Control Measures Distribution by Category}
    \label{fig:category_line}
\end{figure}

In addition, the CCM dataset covers 45 types of capital control measures across key sectors, shown in Figure \ref{fig:category_line}. It includes the capital market, money market, credit operations, derivatives, direct investment, and real estate. Aligned with the classification framework used by \citet{e44f2e54-ab1a-30a8-8fa3-c3bd8f4cef47}, we also provide the measures matched with detailed subcategories such as “Purchase locally by nonresidents” and “Sale or issue abroad by residents.” This level of granularity enables us to infer the directional nature of each measure—whether it targets capital inflows, outflows, or both—based on the specific transaction regulated, and allows us to provide the action direction field in the CCM dataset.

As shown in Figure~\ref{fig:category_line}, most capital control interventions have been implemented in the capital and money market instruments sectors. This pattern is consistent with stylized facts in the literature, which suggest that these sectors are often the first to be targeted during episodes of financial stress.

Notably, the frequency of capital control measures increased sharply around the time of the global financial crisis. Using the event-level CCM dataset, we observe a clear upward trend beginning as early as late 2005, with a steady rise in intervention activity across multiple categories leading up to 2008. A similar pattern emerges around the period of quantitative easing (QE) withdrawal in 2012, when a noticeable increase in capital control announcements followed the Fed’s signaling of QE exit. Similary, from the dataset, we could also observe the increasing capital intervention events during and after the COVID-19 pandemic. These dynamics underscore the value of event-level data in capturing both anticipatory and reactive capital control measures during periods of financial volatility.

\begin{figure}[H]
    \centering
    \includegraphics[width=\textwidth]{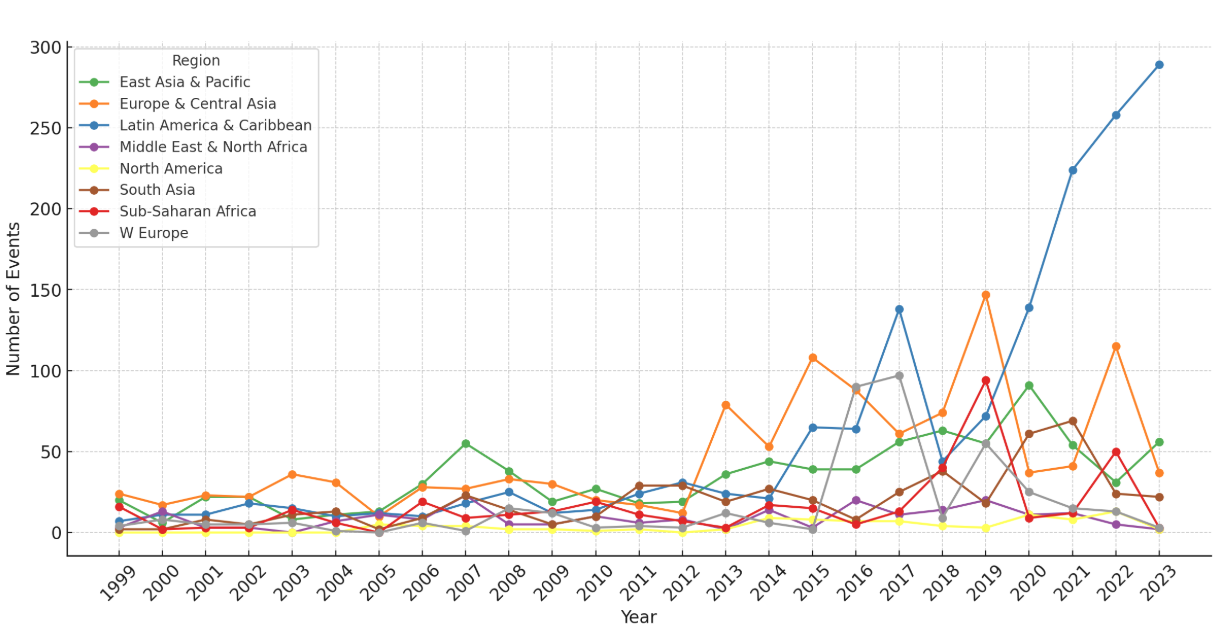}
    \caption{Total Capital Control Measures Distribution by Regions}
    \label{fig:Region}
\end{figure}

In addition to differences by policy category, we also observe substantial regional variation in capital control activity, as shown in Figure~\ref{fig:Region}. The figure displays the time-series evolution of total capital control measures across eight major world regions.

The data indicate that East Asia and Pacific countries responded most actively to the 2008 global financial crisis, while European countries exhibited heightened sensitivity to the monetary tightening associated with the exit from quantitative easing (QE) policies. Notably, capital control activity in Latin America spiked sharply following the COVID-19 outbreak, reaching a record high of approximately 300 interventions in 2023—the largest number of events recorded in a single year for the region.

\begin{figure}[H]
    \centering
    \begin{minipage}[t]{0.8\textwidth}
        \centering
        \includegraphics[width=\textwidth]{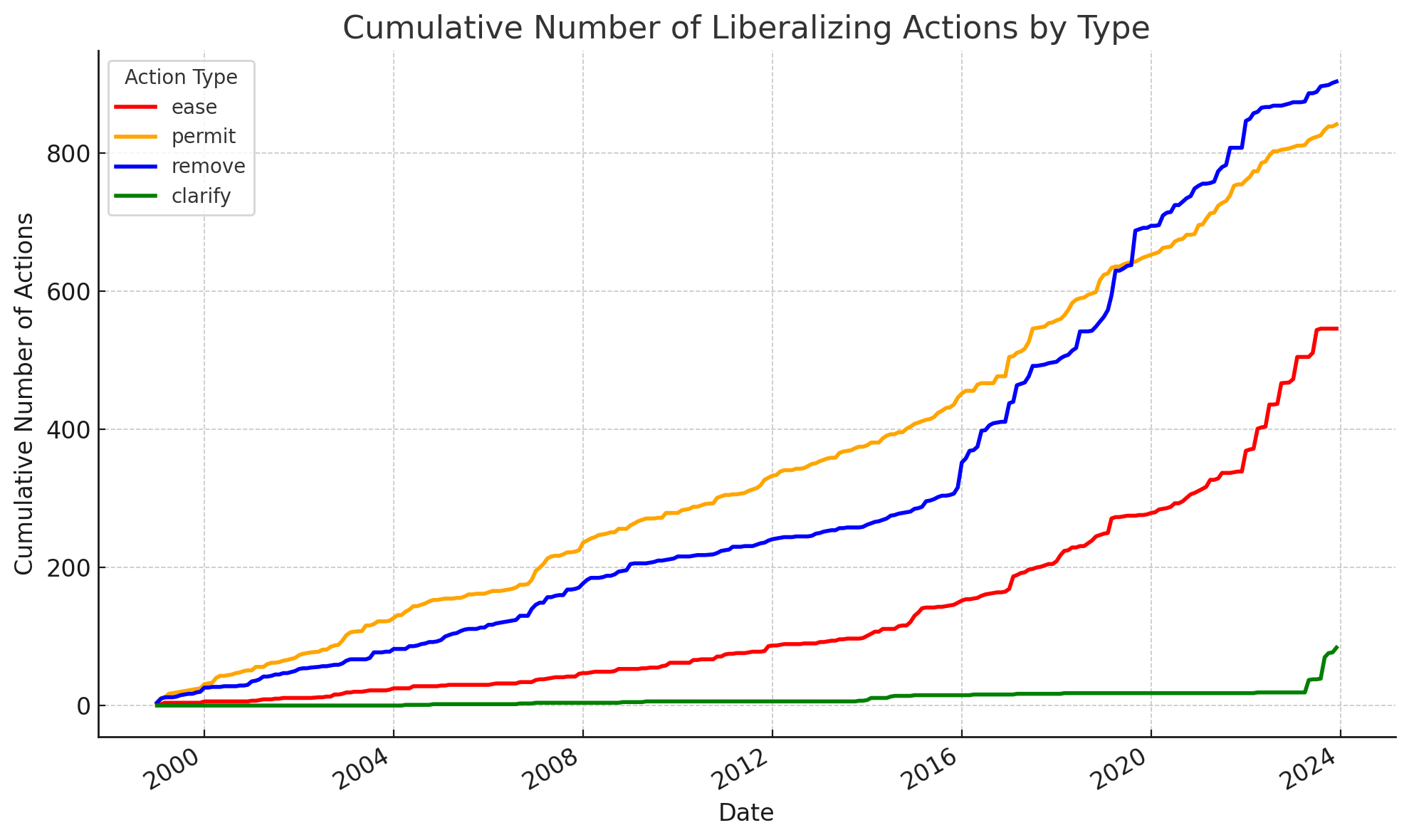}
        \caption*{a. Liberalizing Actions}
        \label{fig:liber_action}
    \end{minipage}
    \vfill
    \begin{minipage}[t]{\textwidth}
        \centering
        \includegraphics[width=0.8\textwidth]{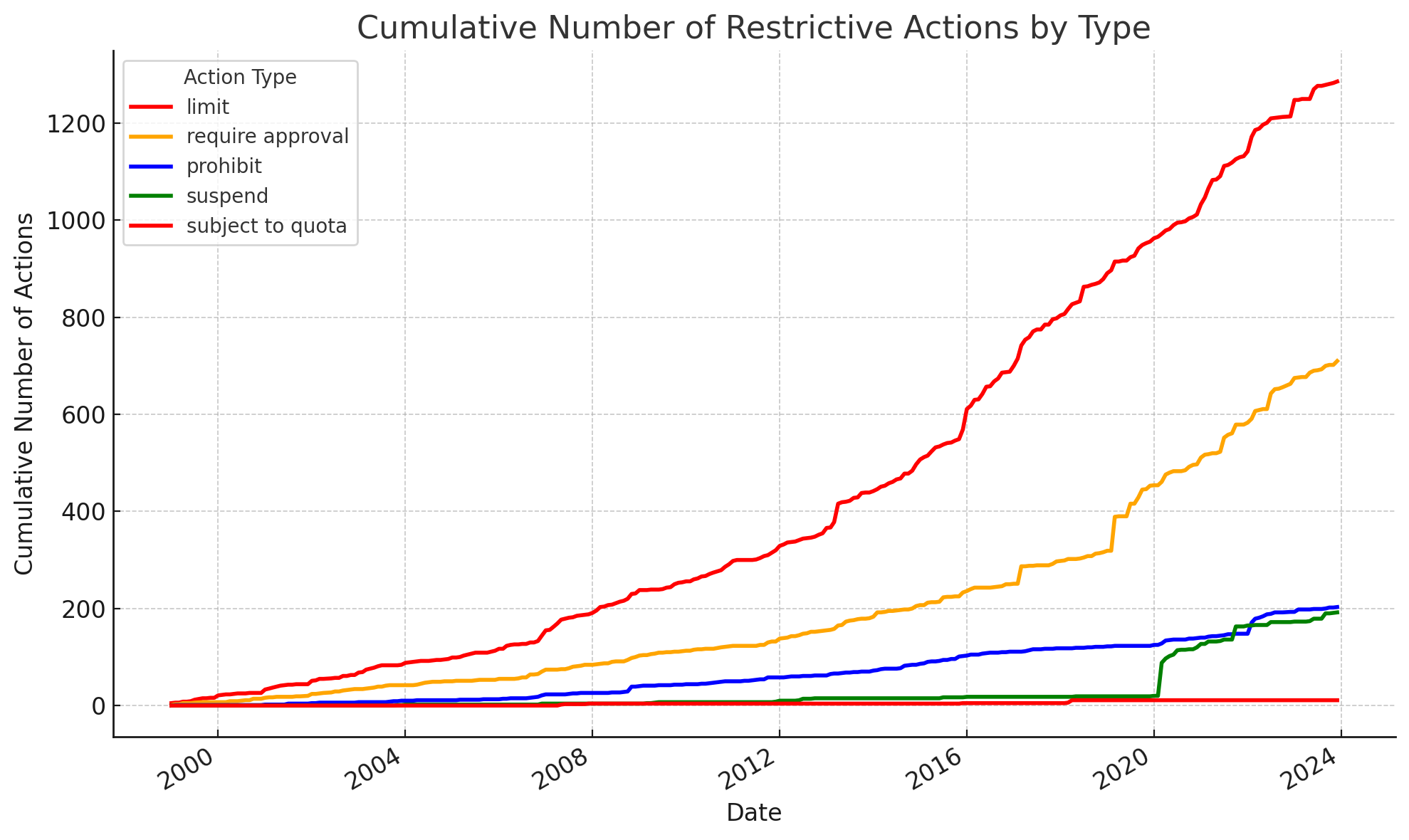}
        \caption*{b. Restrictive Actions}
        \label{fig:restrictive_action}
    \end{minipage}
      \caption{Cumulative CCM by Action Types}
      \label{fig:actions}
\end{figure}

Not only the stylized facts based on the ground-level distribution, we also add more dimensional characteristics based on the rich textual analysis. As we discussed above about the prompting process and the field setup, our dataset introduces a new field by reading the descriptions of each measure and extracting the "action types". The action types help researchers further distinguish the policies with clearer action labels, such as ease, permit, prohibit, or suspend. We provide nine categories of action types, which can be broadly classified into two larger groups:  \textit{Liberalizing Actions} and \textit{Restrictive Actions}. The cumulative frequency of these two types of actions is plotted in Figure~\ref{fig:actions}.

Overall, the number of restrictive actions across all action types is higher than that of liberalizing actions. Within the liberalizing group, as shown in Figure~\ref{fig:actions} part A, “permit” actions appear most frequently, followed by “remove” and “ease” announcements. Beyond the overall trend, it is notable that “remove” actions began increasing more rapidly after 2016 and surpassed “permit” actions around 2020. The number of “ease” actions also spiked following the COVID-19 outbreak, consistent with the notion that governments adopted looser capital control measures to support economic recovery during the post-COVID period. In comparison, Figure~\ref{fig:actions} part B shows that among restrictive measures, “limit” actions consistently outnumber all other types. Similarly, the number of “suspend” actions rose noticeably during the COVID-19 period, indicating tightening in response to heightened uncertainty.

We also classify the action intensity into four types: liberalizing, restrictive, conditional, and neutral. The detailed explanation of how we define and assign these categories is provided in Section~\ref{sec:methodology}. To help the pre-trained LLM understand our classification goals, we provided it with the definitions of each category, along with examples for each type, so that the model had clear references and could better follow our intended classification logic.

\begin{figure}[H]
    \centering

    % 第一行左图
    \begin{minipage}{0.48\textwidth}
        \centering
        \includegraphics[width=\linewidth]{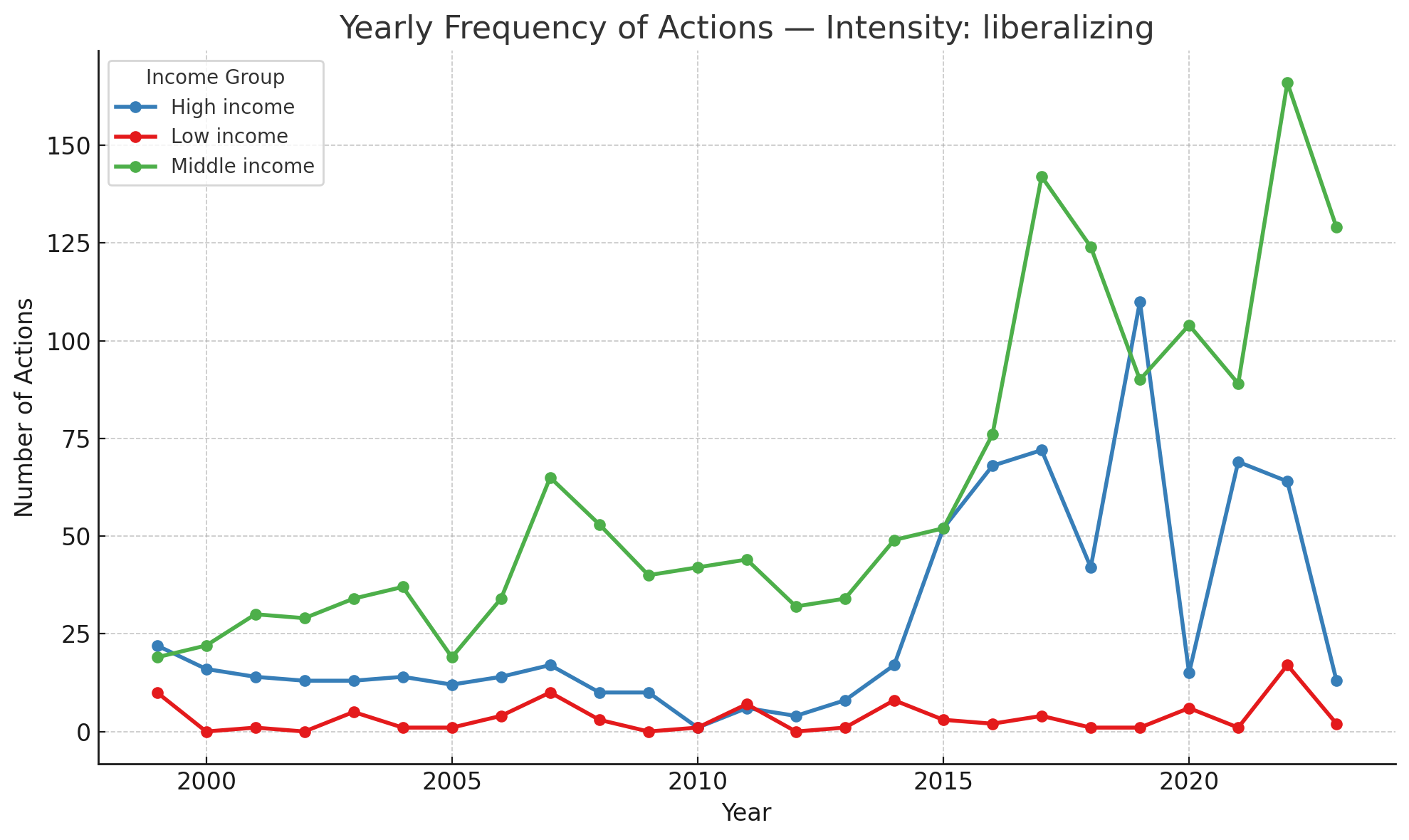}
        \caption*{(a) Liberalizing}
    \end{minipage}
    \hfill
    % 第一行右图
    \begin{minipage}{0.48\textwidth}
        \centering
        \includegraphics[width=\linewidth]{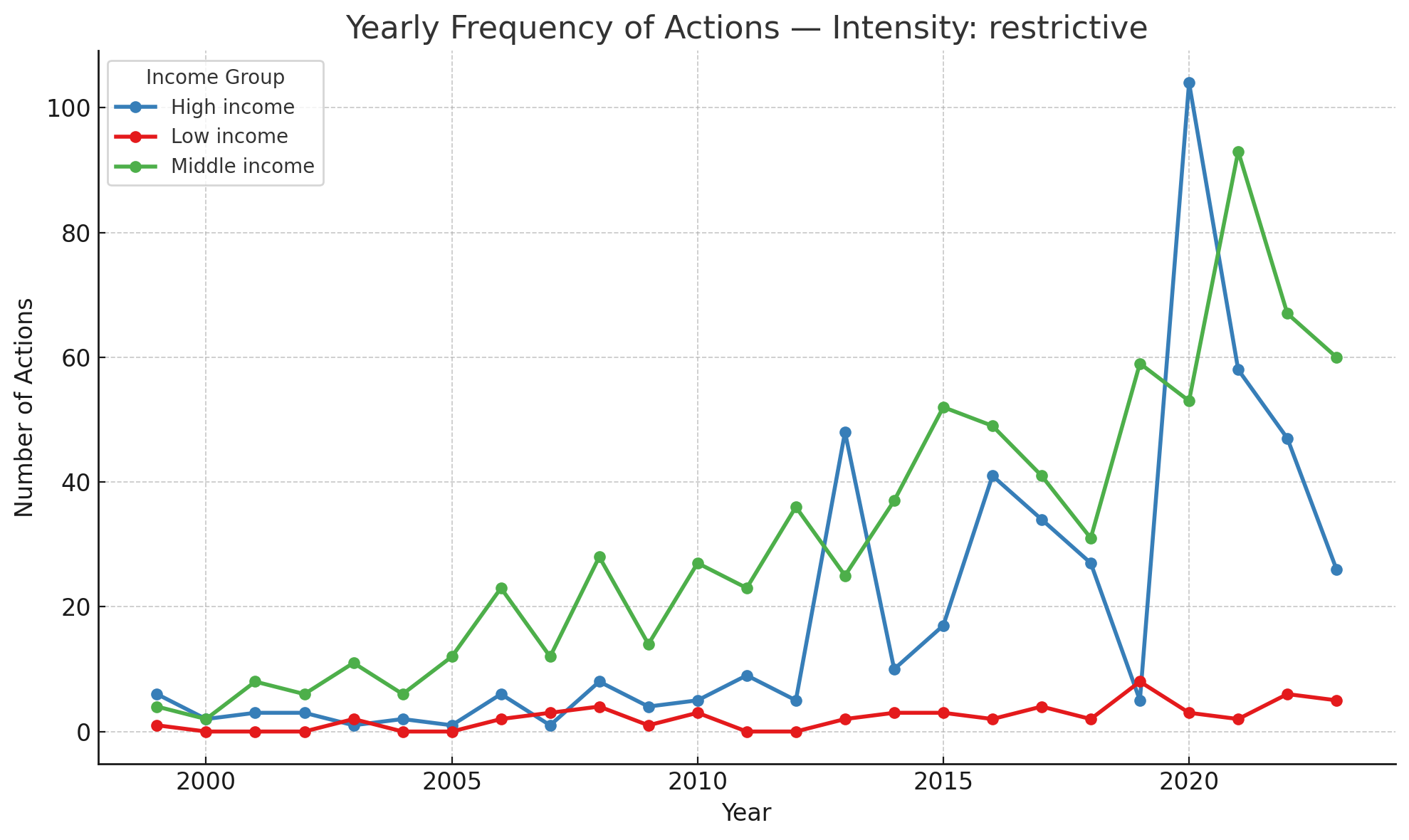}
        \caption*{(b) Restrictive}
    \end{minipage}

    \vspace{1em}

    % 第二行左图
    \begin{minipage}{0.48\textwidth}
        \centering
        \includegraphics[width=\linewidth]{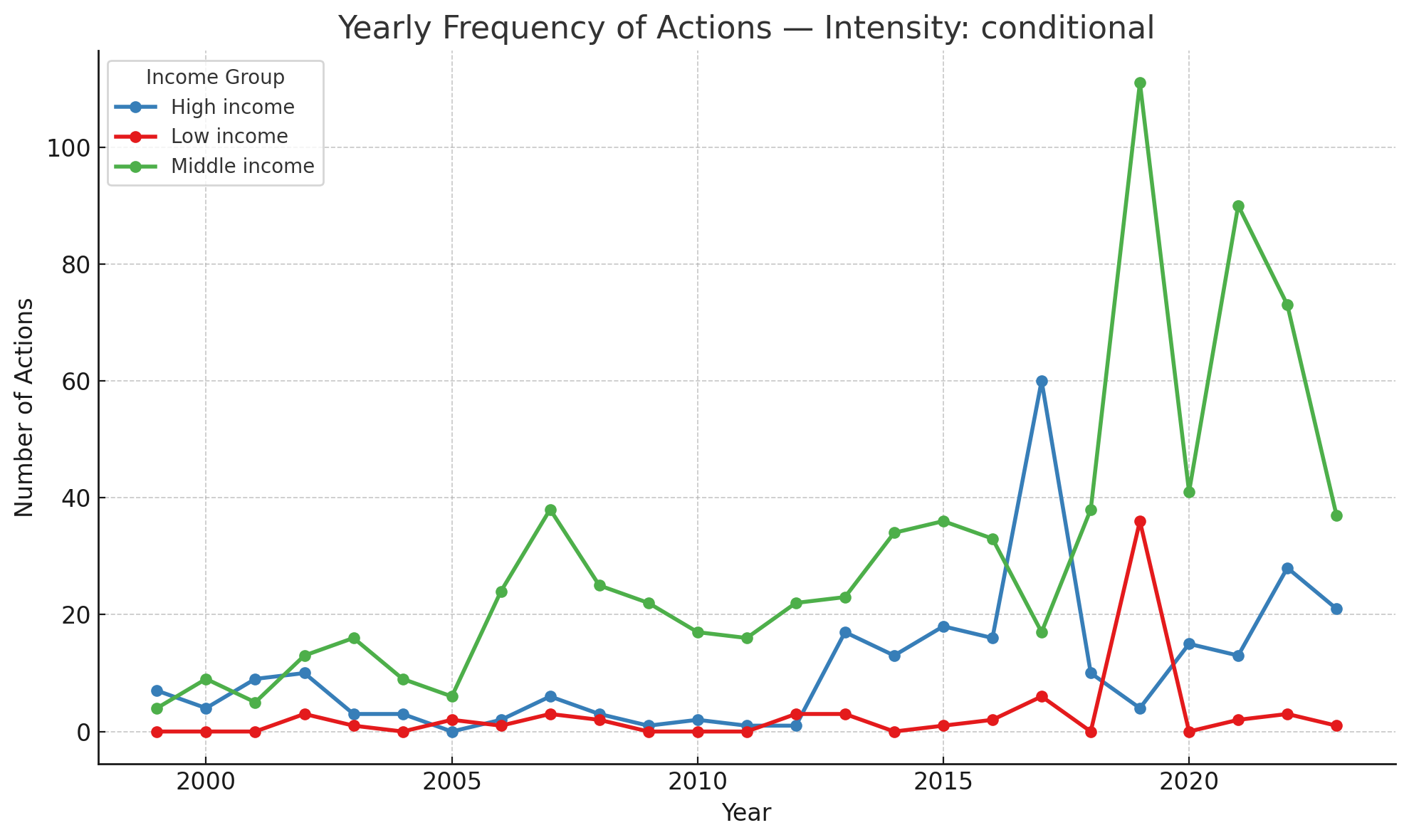}
        \caption*{(c) Conditional}
    \end{minipage}
    \hfill
    % 第二行右图
    \begin{minipage}{0.48\textwidth}
        \centering
        \includegraphics[width=\linewidth]{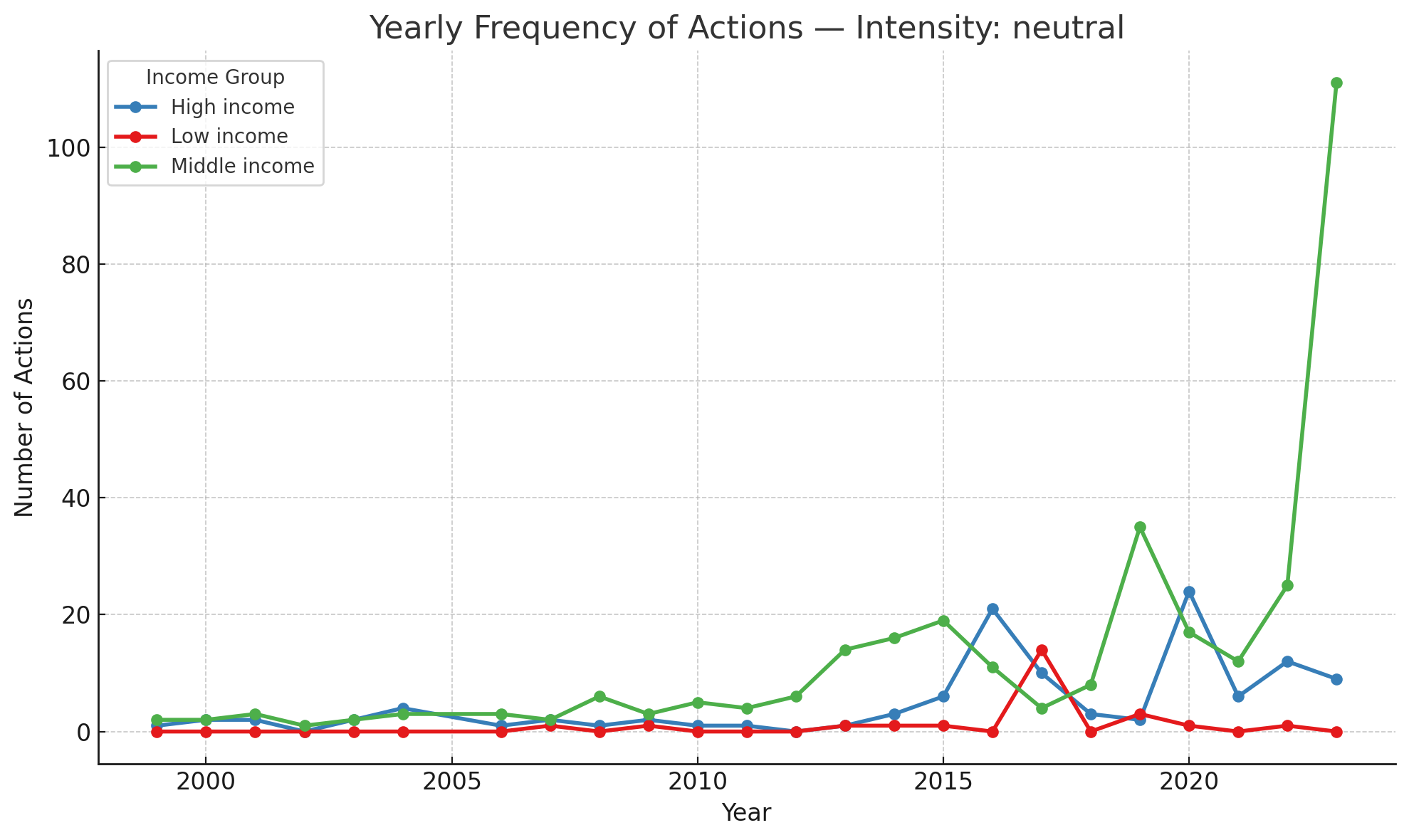}
        \caption*{(d) Neutral}
    \end{minipage}

    \caption{CCM by Action Intensity and Income Group}
    \label{fig:action_intensity}
\end{figure}

Figure~\ref{fig:action_intensity} presents the distribution of capital control measures by income group and action intensity. The four sub-panels show that middle-income countries are, overall, the most frequent users of capital control interventions. A closer look reveals that these countries issued a significant number of liberalizing and conditional measures in the years leading up to the 2008 financial crisis. However, the frequency of such interventions declined sharply during and after the crisis period.

A noteworthy pattern revealed by the action intensity data is that high-income countries tend to lead the trends compared to middle- and low-income groups across all types of capital control actions. This suggests the presence of potential spillover effects, whereby policy shifts in advanced economies may influence the behavior of lower-income countries. The underlying story is that when high-income countries adopt a particular type of capital control, whether imposing restrictions in response to global financial stress or liberalizing to attract capital inflows, other countries often follow with similar interventions shortly thereafter. This pattern may reflect a mix of market contagion, peer effects, and policy learning, as policymakers in emerging and developing economies respond to signals from larger financial centers. While we do not explore this dynamic in detail here, it presents a promising avenue for future research that could investigate the timing, channels, and asymmetry of such cross-country policy transmission.

\begin{figure}[H]
    \centering
    \begin{minipage}[t]{0.7\textwidth}
        \centering
        \includegraphics[width=\textwidth]{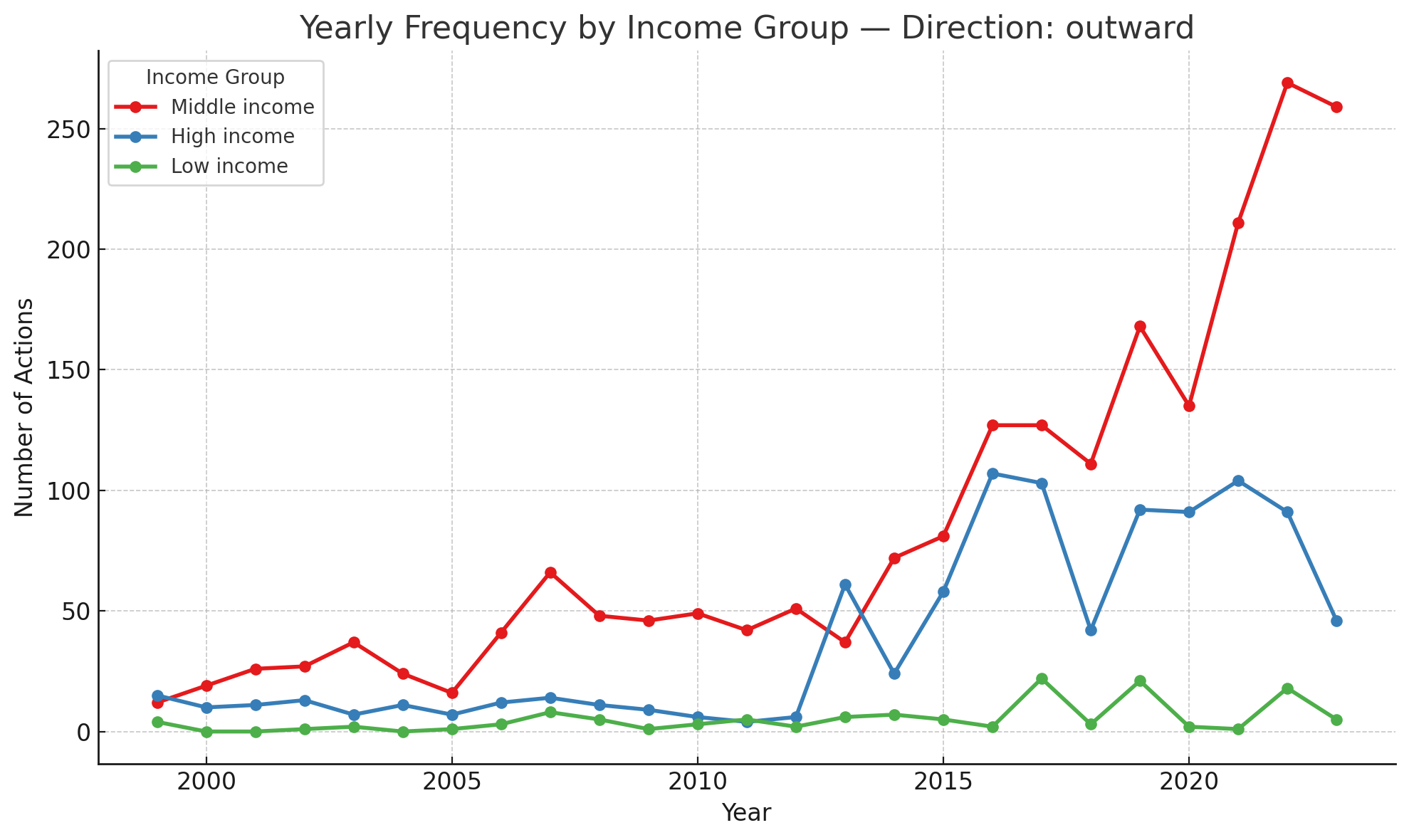}
        \caption*{a. Outward Direction}
        \label{fig:Region1}
    \end{minipage}
    \vfill
    \begin{minipage}[t]{0.7\textwidth}
        \centering
        \includegraphics[width=\textwidth]{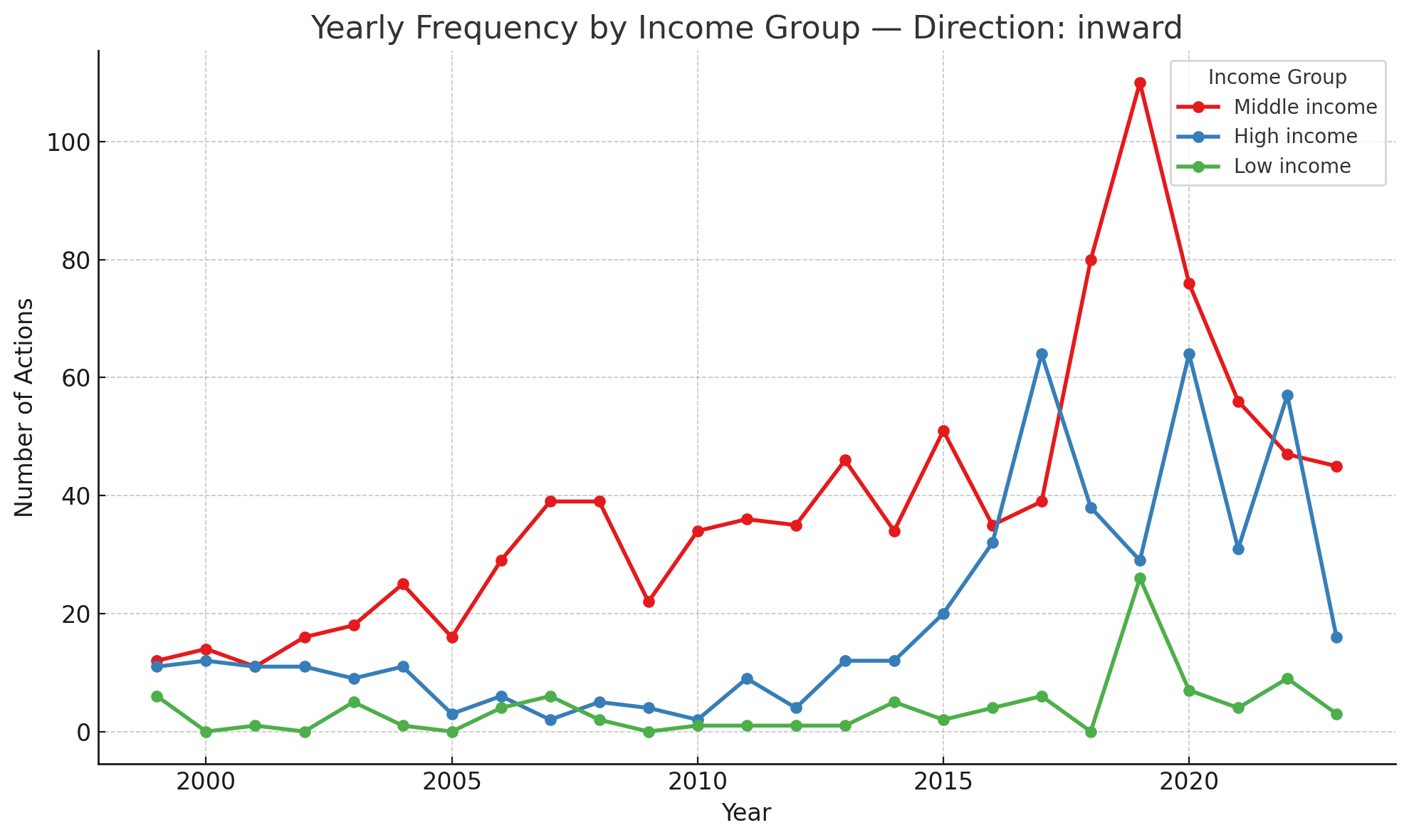}
        \caption*{b. Inward Direction}
        \label{fig:Region2}
    \end{minipage}
    \vfill
    \begin{minipage}[t]{0.7\textwidth}
        \centering
        \includegraphics[width=\textwidth]{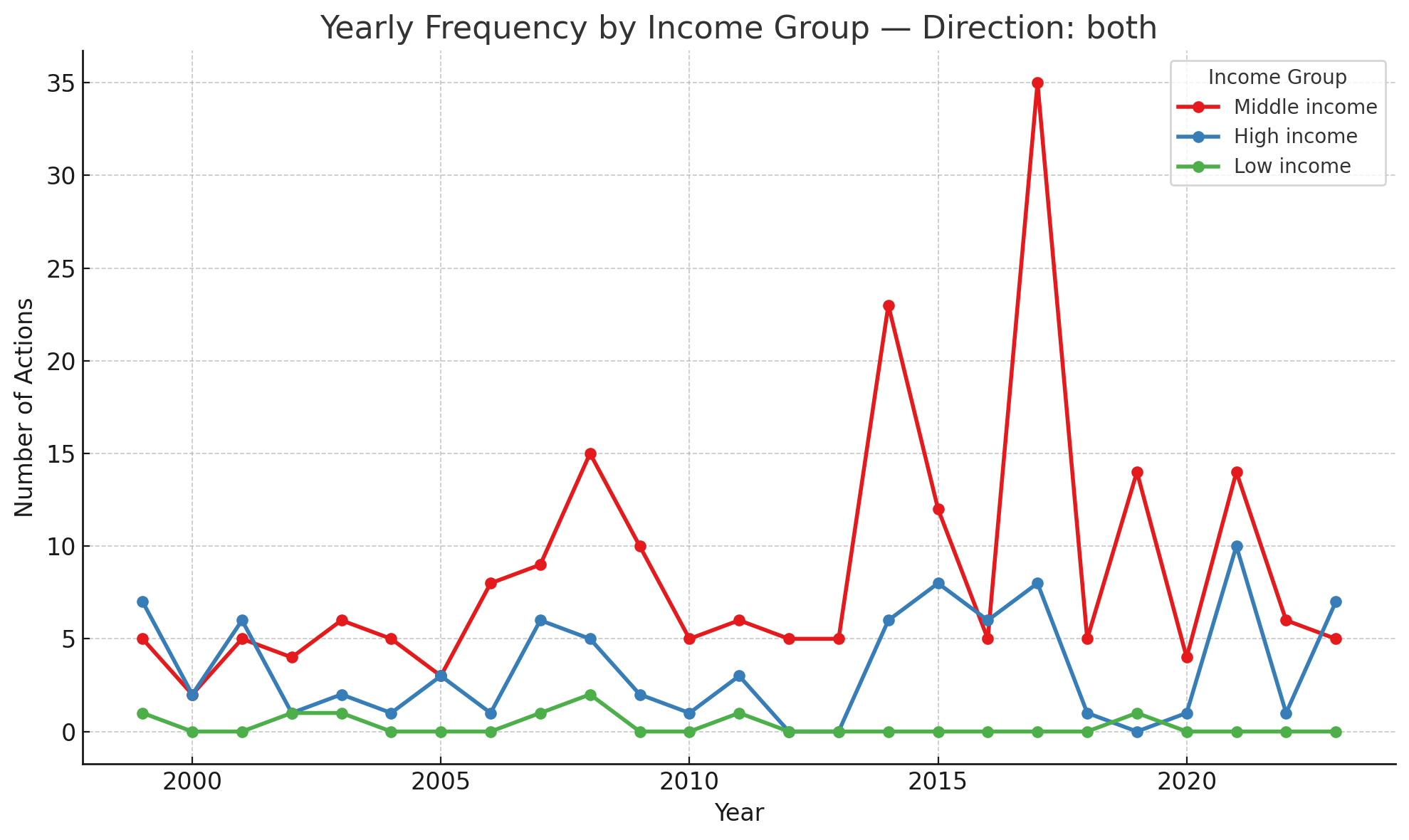}
        \caption*{c. Both}
        \label{fig:Region2}
    \end{minipage}
      \caption{ CCM by Action Directions}
      \label{fig: Action Direction}
\end{figure}

Beyond category-level information that indicates policy directions, the CCM dataset also leverages the textual descriptions in AREAER reports to classify whether each capital control measure targets inflows or outflows. This textual approach provides more robust and interpretable reasoning behind the classification, compared to relying solely on the IMF-assigned categories \footnote{During the data cleaning process, we found that many policy entries fall under broad category labels that lack clear information about whether they apply to capital inflows or outflows. As noted in the technical appendix of Fernández et al.\ (citation), assigning a specific direction often requires manual labeling to match the deepest level of the IMF's classification. This challenge is one of the key motivations for developing our finetuned LLM, described in Section~\ref{sec:finetuning}, which enables automated classification into fine-grained categories using our self-trained open-source \textit{CCM-Llama} 3.1 model.} The figure \ref{fig: Action Direction} is one of the time-series distribution using the \textit{Action Direction} field from the CCM dataset.

In the example shown in the figure above, we continue to classify countries by income level, dividing them into high-, middle-, and low-income groups. It is clear that middle-income countries implemented more outward-related capital control measures than other types of directional actions, and also more than other country groups. By examining these patterns, we can compare how countries responded to the COVID-19 pandemic: in order to prevent capital outflows and manage financial disruptions caused by the crisis, middle-income countries tended to adopt more outward-targeted capital control policies, rather than focusing on inflow restrictions. This is a preliminary observation based purely on the total number of capital control measures. However, with the CCM dataset, we can locate any specific country and category, allowing for deeper analysis of policy performance or reactions in greater detail in future work.

In summary, the CCM dataset captures rich variation in capital control interventions by extracting structured information through LLMs. In this section, we have presented several stylized facts to illustrate the dataset’s advantages and provide a broader sense of its potential uses. Looking ahead, this event-level dataset offers many possibilities for further research across different policy contexts.As an example, in section \ref{sec:empirical}, we conduct a simple empirical analysis using the CCM dataset to implement an event study examining the impact of capital control measures on global fund flows.

\newpage

\section{Classifying Capital Control Categories with Finetuned LLM} \label{sec:finetuning}

In addition to introducing the Capital Control Measures (CCM) dataset, which captures the multidimensional characteristics of each capital intervention policy from the AREAER reports, we further develop an automated categorization process that eliminates human labeling process and is adaptable to multiple data sources. To facilitate the classification of capital control measures and the determination of status conditions for each CCM event, we finetune an open-source large language model based on Meta Llama 3.1~\citep{grattafiori2024llama}, which we named \textit{CCM-Llama}. This model is trained to automatically classify the textual descriptions of capital control measures into the predefined IMF AREAER categories, along with their corresponding status outcomes.

\begin{figure}[!h]
    \centering
    \includegraphics[width=0.95\textwidth]{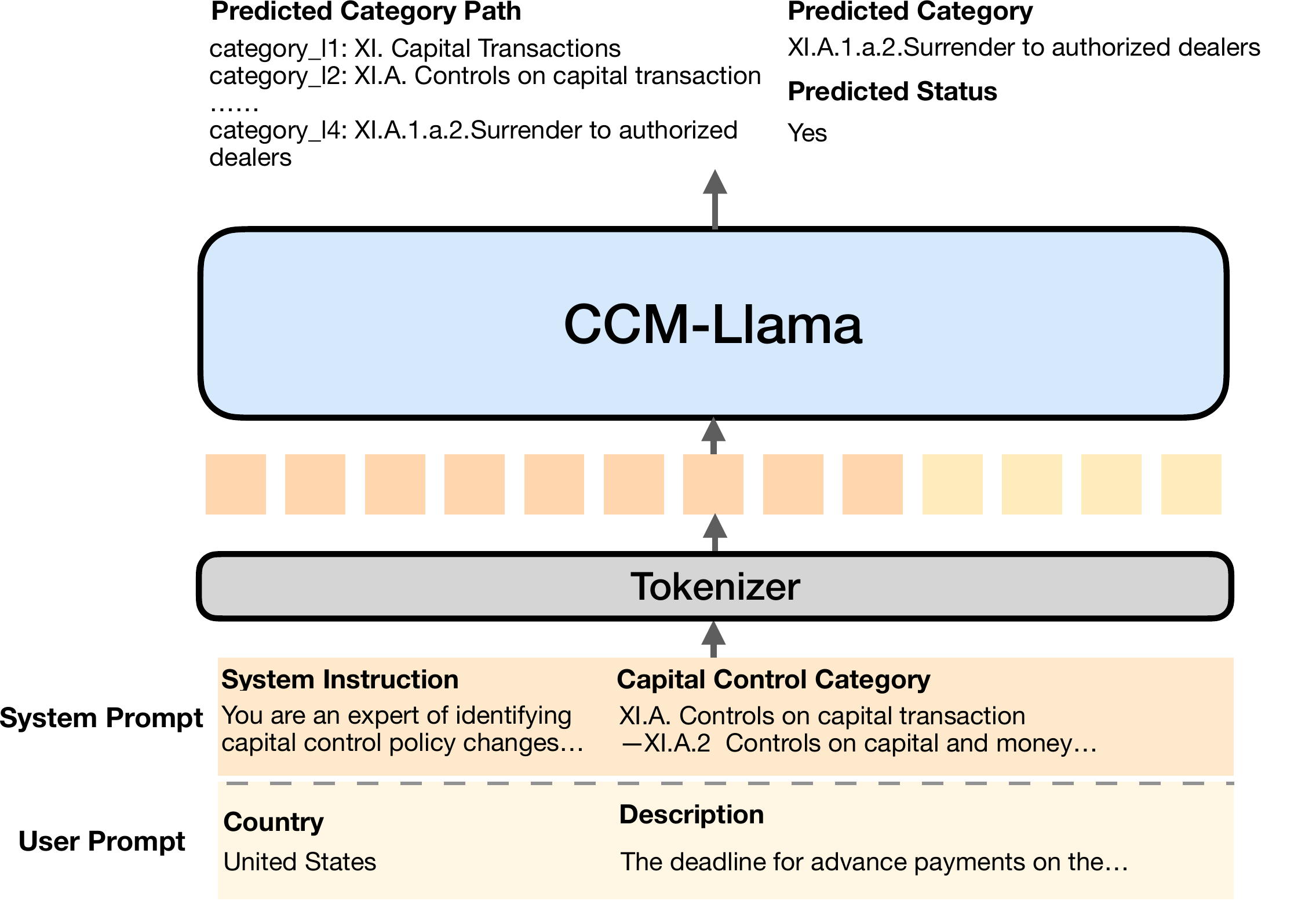}
    \caption{The Inference Process of \textit{CCM-Llama}}
    \caption*{\small \textit{Notes: The model is prompted to classify the description of capital control into IMF AREAER final report categories.}}
    \label{fig:prediction}
\end{figure}

\subsection{Method Overview}

As shown in figure~\ref{fig:prediction}, our goal is to finetune LLMs to understand capital control policy texts and accurately classify them into the relevant categories and statuses defined by the IMF AREAER. 

Specifically, the finetuned LLM is prompted with classification instructions, along with the country name and the textual description of the policy. The input is tokenized into individual tokens to allow for model processing. The model then encodes the tokenized input and outputs the predicted category corresponding to the IMF AREAER.

We design a rigorous three-step pipeline to construct our dataset to enable the finetuning process based on the existing IMF AREAER sources: the final report and the yearly changes. By aligning each policy text with its corresponding category and resulting status, we build a comprehensive training dataset. In total, we collect 29,012 examples covering 196 countries from 1999 to 2022.

With the collected dataset, we finetune the state-of-the-art Meta LlaMA 3.1-8B \citep{grattafiori2024llama} model to make it further understand the relationship between the capital control policy description and category. During finetuning, we apply a standard language modeling approach \citep{zhao2023survey}, training the model to predict the next word in a sequence based on its preceding context. This enables the model to learn how to interpret nuanced policy descriptions and classify them according to IMF-defined categories and status labels. Through this combination of domain-specific data and advanced language modeling, we develop a robust system for automated policy classification. The remainder of this section details our dataset construction steps and finetuning methodology.

% As shown in Figure~\ref{fig:prediction}, our approach involves finetuning LLMs to establish a mapping between textual descriptions of policy events, denoted as $\mathbf{T}$, and categories from the IMF capital control index and their status, represented by $\mathbf{C}$ and $\mathbf{S}$ for a specific country $\mathbf{N}$:

% \begin{equation}
%     \label{eq:llm_mapping}
%     \Tilde{\mathbf{C}},\Tilde{\mathbf{S}}  = \operatorname{LLM}(\mathbf{T}, \mathbf{N}),
% \end{equation}

% \noindent where $\Tilde{\mathbf{C}}$, $\Tilde{\mathbf{S}}$ is the IMF capital control index category and its status inferred by the LLM.

% Using the predicted IMF capital control index category, we subsequently calculate the corresponding predicted capital control index $\mathbf{I}$ following the methodology outlined by {\color{red}CITE(IMF paper)}:

% \begin{equation}
%     \Tilde{\mathbf{I}} = f(\Tilde{\mathbf{C}}, \Tilde{\mathbf{S}}),
% \end{equation}

% \noindenti where $f(\cdot)$ denotes the established mapping rule transforming IMF capital control index categories into numerical capital control indices.

\subsection{Finetuning Dataset Construction}
\label{sec:ftdata_constrc}

Our dataset is based on the IMF AREAER final report and yearly changes. As introduced in Section~\ref{sec:areaer_des}, the AREAER report provides each country's detailed status of each IMF category, while the yearly changes give the published textual policy description and its corresponding changes to a specific IMF category each year.  By combining these two sources, we could extract the input and the ground-truth answer for our training dataset.

Each input consists of: (1) the textual description of a capital control policy, and 2) the name of the country implementing the policy.  We introduce the country information to allow the model to draw on relevant background knowledge stored in its pretrained parameters, which could help disambiguate context-dependent classifications. On the other hand, the ground-truth answer consists of: (1) the IMF category to which the text corresponds, and (2) the resulting status of such category as affected by the described capital control policy.

\begin{figure}[h]
    \centering
    \includegraphics[width=1\textwidth]{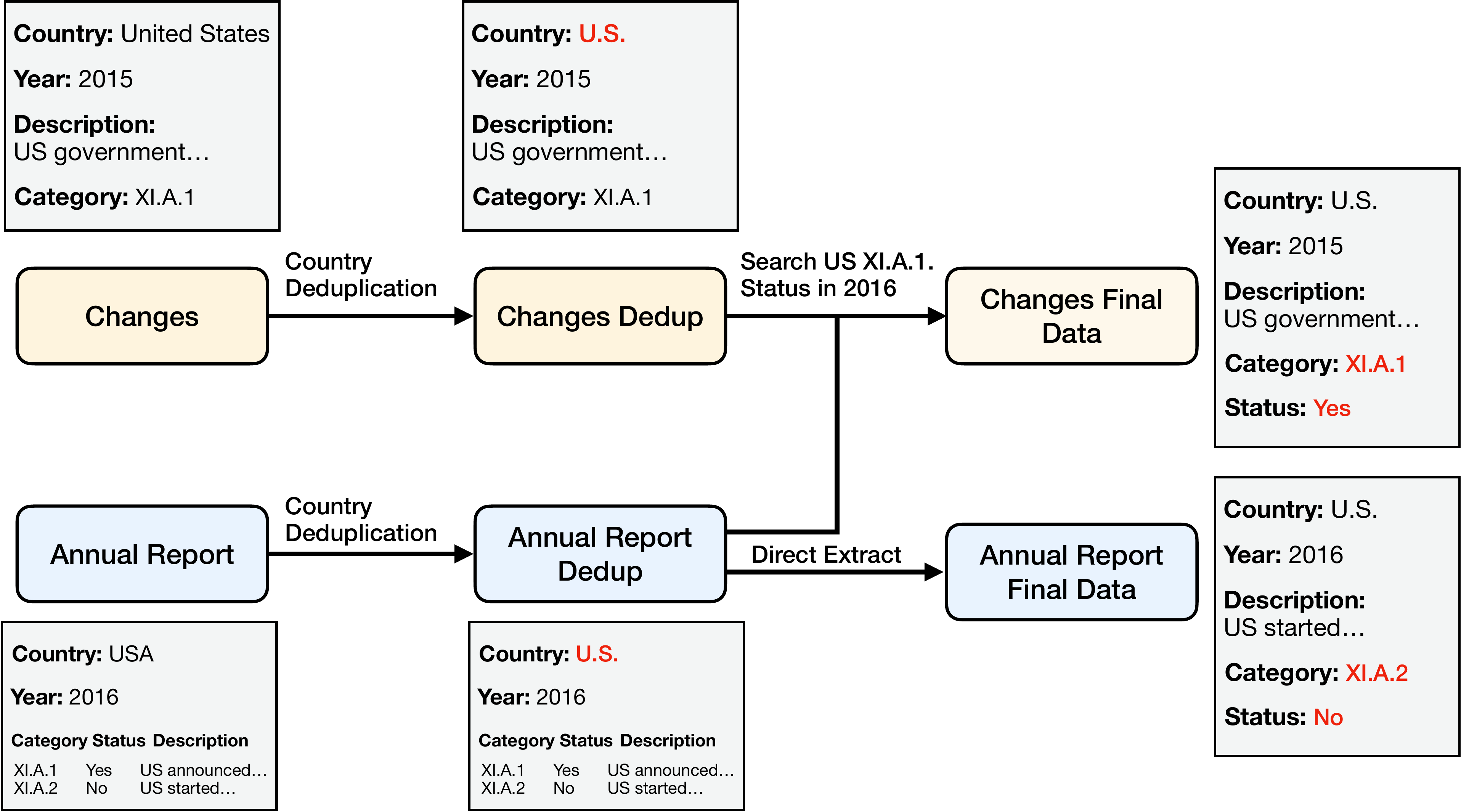}
    \caption{Demonstration of Finetuning Data Collection Process}
    \label{fig:data_collect}
\end{figure}

To robustly construct the inputs and ground-truth labels described above, we design a three-step framework for building our training dataset: (1) Country Deduplication and Data Merging, (2) Training Pair Construction, and (3) Format Transformation. The overall data collection process is illustrated in Figure~\ref{fig:data_collect}.

% These two datasets provide a solid background to support our objective in Equation~\ref{eq:llm_mapping}. 

% Our finetuning dataset is built based on the IMF AREAER annual report where the report includes the textual capital control policy description for XXXXX countries along with their human annotated ``changes" to the capital control index category, 

\noindent\paragraph{Step 1: Country Deduplication and Data Merging} By examining the original report, we identify that some countries in the dataset have multiple representations (e.g., "Hong Kong" and "Hong Kong SAR"), which affects country-level data aggregation. To address this, we first utilize GPT-4o to analyze the unique country names and generate a mapping that merges equivalent representations. A meticulous human review is then conducted to validate the accuracy of GPT-4o’s mapping. The detailed mappings are presented in Table~\ref{tab:country_mapping}. With the standardized country names, we merged all the data from 1999 to 2022 and then grouped them by country for both the final report and yearly changes. \\

\noindent\paragraph{Step 2: Training Pair Construction} 
% At this stage, we transform the merged datasets into the desired input and ground-truth format, $(\mathbf{X}, \mathbf{Y})$. Through careful examination, we identify two types of data points that can be used to form training pairs.

At this stage, we transform the merged datasets into the desired input and ground-truth format. Through careful examination, we identify two types of data that can be used to form training pairs.

The first type is directly derived from the final report. As shown in the lower section of Figure~\ref{fig:data_collect}, for each country in a given year, we collect the textual description of a given IMF category and the country as the input. The corresponding ground-truth answer consists of the category label and its status. This provides naturally aligned input and ground-truth answer pairs.

The second type of data requires the combination of information from both the final report and yearly changes. The process is demonstrated in the upper section of  Figure~\ref{fig:data_collect}.  For a given policy change recorded in the yearly changes for a country $n$ in year $y$, we identify the affected category in the yearly changes. Since the final resulting status is not explicitly written in the yearly changes, we then retrieve the status of the affected category from the final report for country $n$ in year $y + 1$ and regard it as the status caused by the policy change. We then record the input as the policy change with the country information and the ground-truth as the affected category and the final status at year $y+1$. By iterating over all such policy changes, we build a comprehensive and temporally aligned training dataset.

\noindent\paragraph{Step 3: Format Transformation} Since finetuning LLMs requires data to follow the chat completion format\footnote{https://platform.openai.com/docs/guides/text?api-mode=chat}—which includes a system message, a user message, and an assistant message—we perform a format transformation step to adapt our training data accordingly. This step also allows us to incorporate additional background context to help the LLM better understand the task. An exemplar case is shown in Table~\ref{tab:examp_data}.

Specifically, we craft a detailed system message to provide the LLM with domain knowledge and task-specific instructions. This message includes all relevant category options related to capital controls, and provides: (1) the hierarchical structure of the categories based on IMF AREAER final report, and (2) a concise one-sentence description for each category. The detailed hierarchy and description are shown in Table~\ref{tab:cc_des} in Appendix~\ref{app:cc_des}. This enriched context helps the LLM more accurately understand both the domain and the classification choices \citep{liu-etal-2024-beyond-text}. The user message, which serves as the query in the chat format, includes the country and the policy change text. The assistant message, representing the LLM's expected response, follows a structured reasoning schema. The outputs of the assistant message contains: (1) the reasoning path leading to the final answer, guided by the category hierarchy defined in the system message; (2) the inferred change category; (3) the updated status.

\begin{table}[!t]
\centering
\caption{Exemplar Chat Format for Finetuning Dataset}
\begin{tabular}{@{}p{2.5cm}p{12cm}@{}}
\toprule
\textbf{Message} & \textbf{Content} \\
\midrule
\textbf{System} & You are an expert in identifying the capital policy change and updating the IMF category status based on the input text. Please select from outer layer to the inner layer. Please consider both the country's capital control status and background and the given text. Here are the hierarchical structures for the category with its description: \par

\textbf{XI.A.} Controls on capital transactions — Restrictions on cross-border capital flows, covering investment, credit, and real estate transactions. \par
- \textbf{XI.A.2.} Controls on capital and money market instruments — Rules regulating international transactions in equity, debt, money markets, and investment funds. \par
\hspace{1em}- \textbf{XI.A.2.a.} On capital market securities — Measures on shares and bonds issued or acquired by residents or nonresidents. \par
\textellipsis{} (truncated for brevity) \\
\midrule
\textbf{User} & \textbf{Country:} United States \newline
\textbf{Policy change:} Executive Order 13722, \textit{Blocking Property of the Government of North Korea and the Workers’ Party of Korea and Prohibiting Certain Transactions with respect to the Democratic People’s Republic of Korea}, was issued. \\
\midrule
\textbf{Assistant} & \texttt{\{"category\_path": \{"category\_l1": "XI.Capital Transactions", "category\_l2": "XI.A.Controls on capital transactions", "category\_l3": "XI.A.5.Controls on direct investment", "target\_category": "XI.A.5.b.Inward direct investment"\}, "index": "XI.A.5.b.", "category": "Inward direct investment", "status": "yes"\}} \\
\bottomrule
\end{tabular}
\label{tab:examp_data}
\end{table}

In addition, to make sure the model could distinguish capital control measures from unrelated policy measures, we also include the non-capital-control sections of the final report and yearly changes into the training dataset, with the same data collection and formatting procedure applied.

\begin{figure}[!h]
    \centering
    \includegraphics[width=\textwidth]{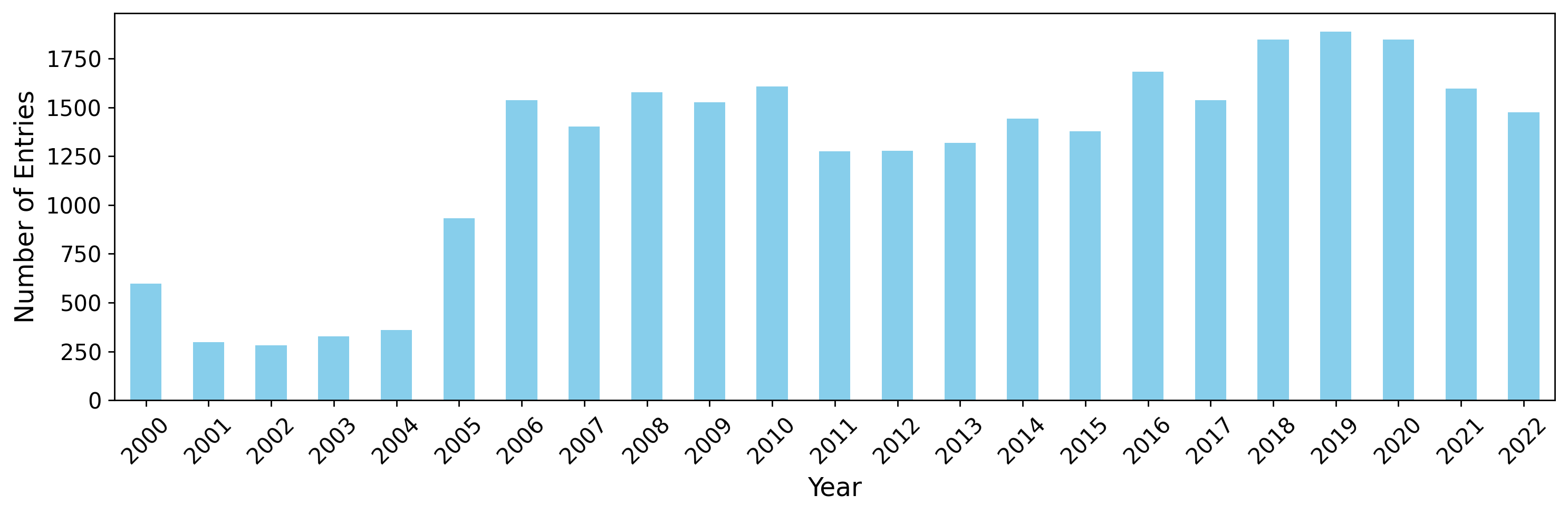}
    \vspace{-1.5em}
    
    \includegraphics[width=\textwidth]{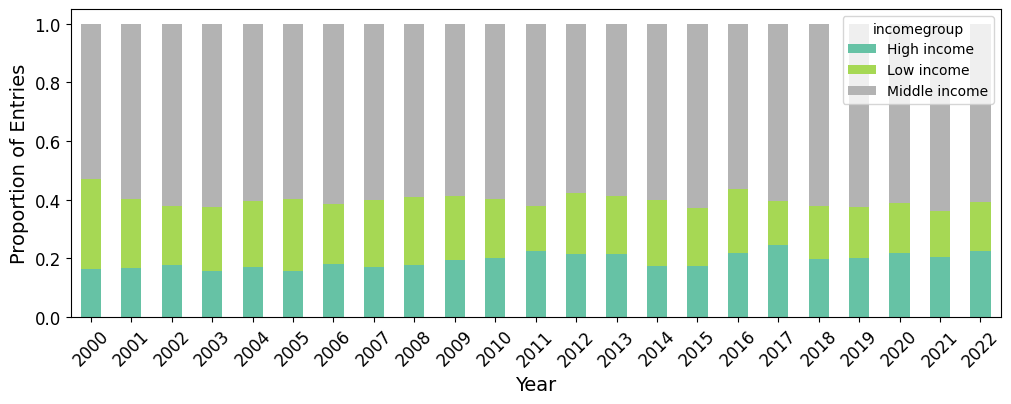}
    \vspace{-1.5em}
    
    \includegraphics[width=\textwidth]{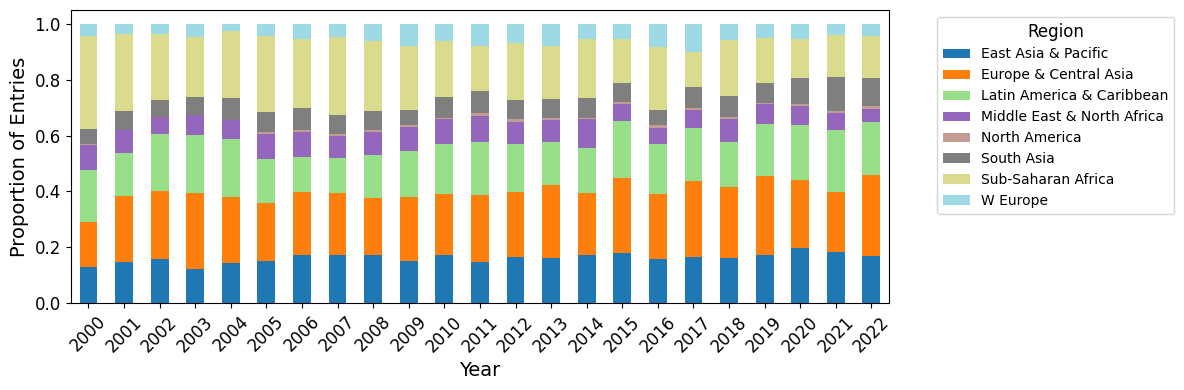}
    
    \caption{Distribution of Our Finetuning Dataset}
    \caption*{\small \textit{Top: Data distribution by year; Middle: Data distribution by income group; Bottom:  Data distribution by region}}
    \label{fig:traindata_combined_dist}
\end{figure}

\begin{figure}[!h]
    \centering
    \includegraphics[width=\textwidth]{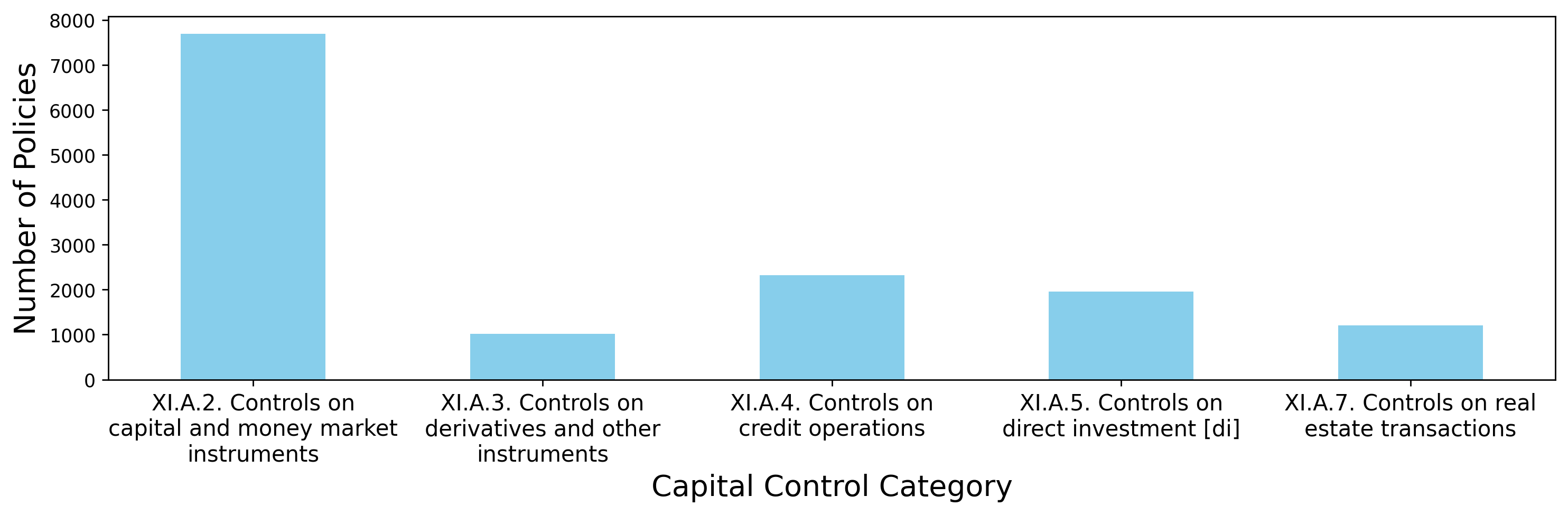}
    \vspace{-1.5em}
    
    \includegraphics[width=\textwidth]{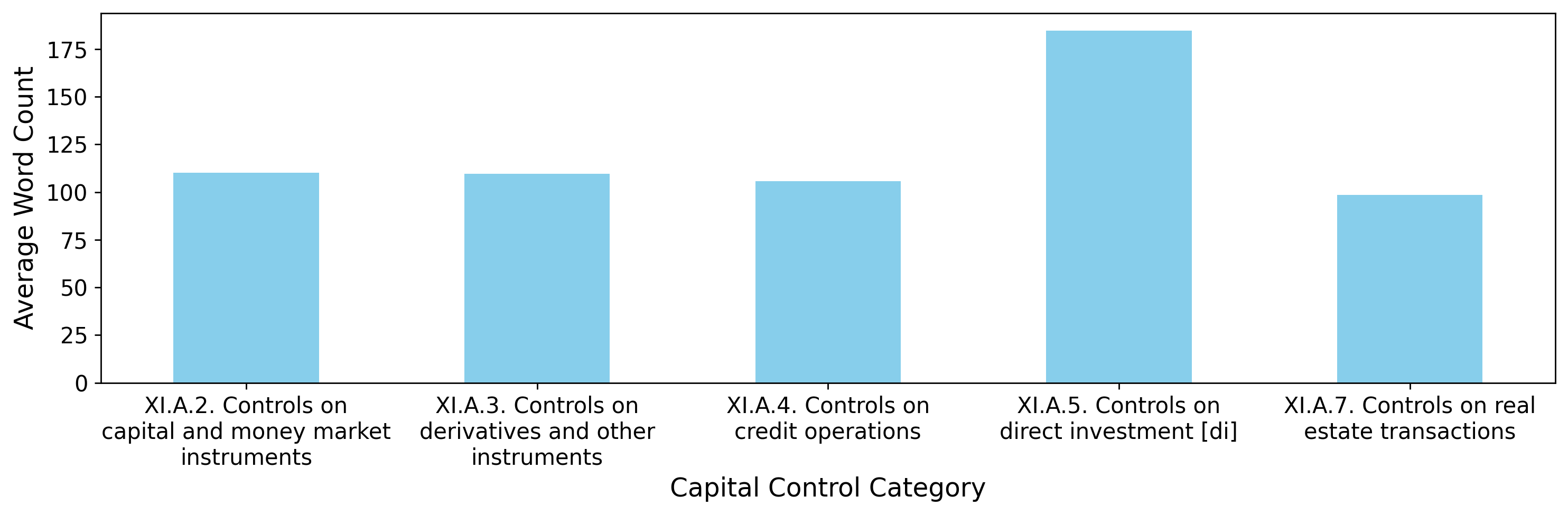}
    
    \caption{Distribution of Capital Control Data}
    \caption*{\small \textit{Top: Distribution of Capital Control Data by Category;  Bottom: The Average Word Count of Each Category.}}
    \label{fig:traindata_cc_dist}
\end{figure}

\noindent\paragraph{Finetuning-Dataset Descriptions}
In total, our dataset consists of 29,012 examples spanning 196 countries from 1999 to 2022. In detail, 14,182 examples fall within capital control categories, while the remaining 14,830 belong to other policy categories. This balanced distributions cross the related categories and unrelated ones enhance our model to have the capability of learning distinguishment effectively. 

Additionally, we analyze the dataset distribution by year, income group, and region. As shown in Figure~\ref{fig:traindata_combined_dist}, our dataset covers all income groups and regions. We ensure a balanced representation across years and groups to better support effective model learning. Finally, we closely examine the capital control policy data. As presented in Figure~\ref{fig:traindata_cc_dist}, our dataset includes examples from every capital control category defined by the IMF. Moreover, the average text length exceeds 100 words, ensuring that each policy description provides sufficient contextual information for accurate classification.

% \begin{figure}[!t]
%     \centering
%     \includegraphics[width=\textwidth]{fig/dist_year.pdf}
%     \caption{Distribution of Year (Appendix)}
%     \label{fig:dist_year}
% \end{figure}

% \begin{figure}[!t]
%     \centering
%     \includegraphics[width=\textwidth]{fig/dist_countries.pdf}
%     \caption{Distribution of Country (Appendix)}
%     \label{fig:dist_country}
% \end{figure}

% \begin{figure}[!t]
%     \centering
%     \includegraphics[width=\textwidth]{fig/dist_cate.pdf}
%     \caption{Distribution of Category (Appendix)}
%     \label{fig:dist_cate}
% \end{figure}

\subsection{Model Finetuning Process}

\begin{figure}[!t]
    \centering
    \includegraphics[width=0.85\textwidth]{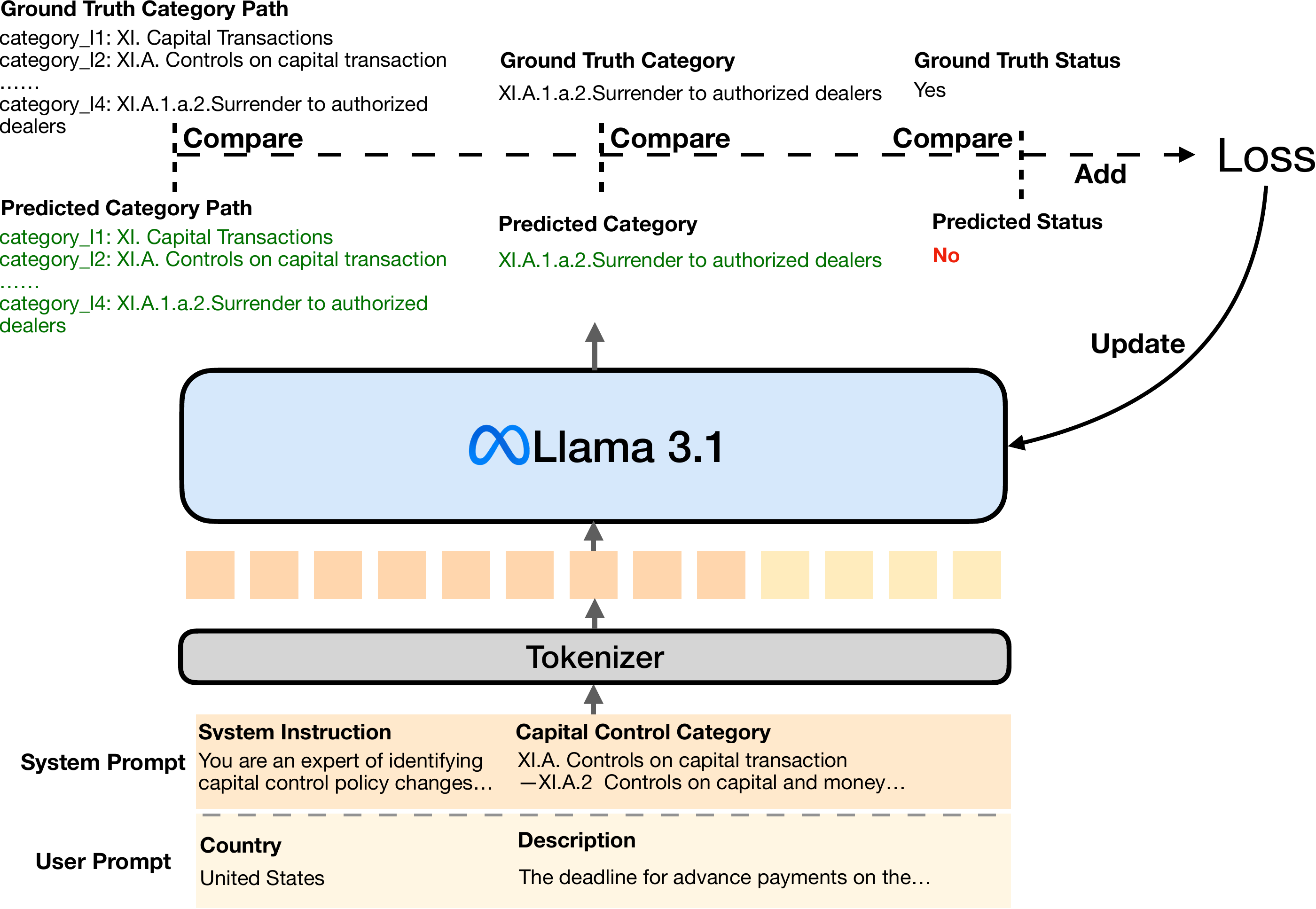}
    \caption{The Model Finetuning Process}
    \label{fig:training}
\end{figure}
\noindent\paragraph{Implementation Details} We split our dataset of 29,012 examples into training, validation, and test sets following a standardized procedure. The training set contains 28,412 examples, while the validation and test sets contain 100 and 500 examples, respectively.
Model training is conducted on two NVIDIA A6000 GPUs. We use a learning rate of $1 \times 10^{-5}$ with a linear decay schedule, a batch size of 1, and evaluate the model every 200 training steps using the validation set. The final model is selected based on its best performance on the validation data. Training is run for a total of 2 epochs.

We leverage Meta’s LLaMA 3.1 \citep{grattafiori2024llama} as our pretrained base model due to its extensive training on diverse and high-quality textual corpora, including substantial content from economics and finance domains. This broad coverage provides a strong foundation for downstream adaptation. We initialize our model using the official checkpoint available via Hugging Face\footnote{\url{https://huggingface.co/meta-llama/Llama-3.1-8B}} and finetune the entire model to maximize its adaptability to our specific task.

\noindent\paragraph{Finetuning Objective}  Our finetuning procedure follows the standard language modeling paradigm. For each data point we obtain from Section \ref{sec:ftdata_constrc}, we first use the model's tokenizer to break the full text into a sequence of tokens $t = (t_1, t_2, \ldots, t_n)$. Then the model is trained to predict the next token at each position. Specifically, at position $i$, the model receives the prefix sequence $(t_1, \ldots, t_{i-1})$ as input and learns to predict the token $t_i$. The training objective minimizes the negative log-likelihood loss over the entire sequence:
\begin{equation}
    \mathcal{L}_{\text{LM}} = - \sum_{i=1}^{n} \log P_\theta(t_i \mid t_1, \ldots, t_{i-1}),
\end{equation}

where $P_\theta$ denotes the probability distribution over the vocabulary parameterized by the model weights $\theta$. This autoregressive loss encourages the model to generate coherent and contextually appropriate text by learning the conditional distributions of tokens within the training corpus. The finetuning process is shown in Figure ~\ref{fig:training}.

\subsection{Finetuning Experiment Result}

To evaluate the effectiveness of our finetuned model, we conduct comprehensive performance comparisons between our finetuned \textit{CCM-Llama} and a range of state-of-the-art commercial and open-source language models.

\noindent\paragraph{Baseline Models} We compare \textit{CCM-Llama} against leading foundation models from both commercial providers and the open-source community. For commercial models, we include OpenAI GPT-4o \citep{gpt4o2024} and GPT-4.1 \citep{gpt412025}, which represent the current state-of-the-art. For open-source models, we evaluate against Meta LLaMA 3.1-8B \citep{grattafiori2024llama}, enabling a direct comparison with our finetuned counterpart.

\noindent\paragraph{Benchmark Dataset} Our experiment is conducted on the test split of our dataset which contains 500 examples. The dataset contains 234 capital control-related policies and 266 non-capital control policies.

\noindent\paragraph{Evaluation Metrics} We evaluate our model based on two key questions: (1) Can the model accurately distinguish between capital control and non-capital control policies? (2) For policies identified as capital controls, can the model correctly classify them into the appropriate sub-category?

To address the first question, we compute the binary classification accuracy, defined as:

\begin{equation}
    \text{Accuracy}_{\text{binary}} = \frac{1}{N} \sum_{i=1}^{N} \mathbb{I}(y_i = \hat{y}_i)
\end{equation}

where \( y_i \in \{0,1\} \) is the ground-truth label (1 for capital control, 0 for non-capital control), \( \hat{y}_i \) is the model's prediction, \( N \) is the total number of samples, and \( \mathbb{I}(\cdot) \) is the indicator function that returns 1 if the argument is true and 0 otherwise. \\

To evaluate the second question, we compute hierarchical accuracy across six levels of the classification hierarchy (denoted as L1 through L6). Each category is represented by a hierarchical index (e.g., \texttt{XI.A.2.a.1.ii}), which we decompose into up to six levels using dot-separated components.

For each policy where the ground-truth is a capital control policy, we compare its predicted index to the ground-truth index level by level. A match at level \( L_k \) is counted only if \textit{all previous levels} (1 to \( k-1 \)) also match. Once a mismatch occurs at any level, deeper levels are not counted as matches. This enforces a strict top-down matching structure.

Formally, for each level \( L_k \in \{1, 2, \ldots, 6\} \), we define:

\begin{equation}
    \text{Accuracy}_{L_k} = \frac{\text{\# predictions where levels 1 through }k\text{ match}}{\text{\# instances with valid ground truth at level }k}
\end{equation}

\begin{table}[!t]
\centering
\caption{Experiment results for capital control binary and hierarchical classification}
\caption*{\small \textit{\textbf{Binary Acc} represents binary accuracy, while \textbf{Hierarchical Acc} represents accuracy at each hierarchy level. \textbf{$\Delta$ over Best Baseline} indicates the percentage point improvement of CCM-Llama over the best-performed baseline model.}}
\label{tab:ft_exp_result}
\begin{tabular}{l|c|ccccc}
\toprule
\textbf{Model} & \multirow{2}{*}{\textbf{Binary Acc}} & \multicolumn{5}{c}{\textbf{Hierarchical Acc}} \\
\cmidrule(lr){3-7}
              &                           & \textbf{L3} & \textbf{L4} & \textbf{L5} & \textbf{L6} & \textbf{Avg} \\
\midrule
Llama 3.1-8B   & 90.67                  & 84.00 & 59.05 & 44.68 & 26.42 & 53.54 \\
GPT-4o        & 76.07                  & 70.09 & 59.63 & 42.07 & 40.74 & 53.13 \\
GPT-4.1       & 94.44                  & 82.91 & 68.81 & 56.55 & 48.15 & 64.10 \\
\midrule
CCM-Llama     & \textbf{99.55}         & \textbf{90.09} & \textbf{75.85} & \textbf{65.19} & \textbf{60.78} & \textbf{72.98} \\
\rowcolor{gray!10}
$\Delta$ over Best Baseline& \textbf{+5.11}  & \textbf{+7.18} & \textbf{+7.04} & \textbf{+8.64} & \textbf{+12.63} & \textbf{+8.88} \\
\bottomrule
\end{tabular}
\end{table}

\noindent\paragraph{Results and Analysis}

As shown in Table~\ref{tab:ft_exp_result}, \textit{CCM-Llama} achieves the highest binary classification accuracy at \textbf{99.55\%}, substantially outperforming all the baseline models. This demonstrates that our model is highly reliable at identifying whether a given policy falls under the category of capital controls. 

For the hierarchical accuracy evaluation, since capital control categories all start with \texttt{XI.A}, we report the accuracy from L3 to L6. In general, performance degrades with deeper levels, reflecting the increasing difficulty of fine-grained classification. Nevertheless, \textit{CCM-Llama} consistently outperforms all baselines at each level. Especially at more fine-grained levels, L5 and L6, \textit{CCM-Llama} maintains strong performance with more than 60\% accuracy while other models drop more steeply down to less than 50\%.

Overall, \textit{CCM-Llama} achieves the highest per-layer accuracy, reaching up to \textbf{90.09\%}, and an average sub-category accuracy of \textbf{72.98\%} across levels L3 to L6. These results significantly outperform the best baseline model by up to 12.63\%. This demonstrates the effectiveness of our domain-specific finetuning approach for structured policy classification tasks, especially in capturing subtle and complex hierarchical distinctions.

\noindent\paragraph{Discussion} The results underscore the limitations of general-purpose LLMs, even state-of-the-art commercial models, on specialized classification tasks involving capital control policy texts. Despite their scale and general reasoning capabilities, these models underperform relative to a targeted, finetuned model like \textit{CCM-Llama}, especially in fine-grained, taxonomy-aware prediction. These findings support the case for domain adaptation when applying LLMs to high-stakes policy domains.

With its robust performance, \textit{CCM-Llama} unlocks broad potential for the automatic real-time and multi-source detection of capital control events. Unlike prior approaches that rely solely on human beings and gathering interventions from AREAER reports, our model can be deployed across diverse text sources, including financial news corpora such as Refinitiv News Portal and Bloo-mberg News Updates, which offer richer contextual information and higher temporal granularity. This flexibility opens the door to a wide range of applications, such as near real-time monitoring of policy interventions, early-warning systems for capital flow volatility, and enhanced input for empirical macro-finance research where timely policy signals are critical.

\section{Capital Control Measures on Global Fund Flow: An Empirical Analysis} \label{sec:empirical}
One of the key advantages of the CCM dataset is its event-based structure, which allows economists to analyze the effects of specific capital control actions with high temporal precision. In this section, after constructing the dataset through prompt-based analysis of the AREAER reports, we show an empirical example to use the CCM dataset to examine how different types of policy actions, and their action intensity, affect global fund flows.
\subsection{Datasets}
\noindent\paragraph{Fund Flow Dataset} Our dependent variable is constructed using the EPFR Global fund flow dataset, which tracks capital flows from mutual funds, ETFs, and other portfolio investment vehicles across countries. The dataset covers 47 countries from 2008 to 2017 at a monthly frequency and provides detailed information on how global fund assets are allocated across destination markets.

Using the fund flow dataset, we are able to identify both the absolute inflow amount and the relative share of capital directed toward each geographically focused country. We construct the fund flow percentage based on the EPFR Global dataset. The first measure is the absolute fund flow into country $i$ at time $t$, defined as:

\[
\text{Fund Flow}_{i,t} = \sum_j \text{Fund Size}_{j,t} \times \text{Weight}_{j,i,t}
\]

where $\text{Fund Size}_{j,t}$ denotes the total net asset value of the fund $j$ at time $t$, and $\text{Weight}_{j,i,t}$ represents the share of the fund $j$’s investment allocated to country $i$. To account for differences in exposure across countries, we normalize this variable by the total fund size allocated to country $i$ at time $t$:

\[
\text{Fund Flow Percentage}_{i,t} = \frac{\text{Fund Flow}_{i,t}}{\text{Total Fund Size}_{i,t}}
\]

This measure captures the proportion of capital allocated to country~$i$ relative to the total size of investment positions directed toward that country. In future research, these measurements can also be extended to compare different types of fund flows: such as ETFs, mutual funds, and other investment vehicles, based on their distinct dynamics and sensitivities.

\noindent\paragraph{Capital Control Measures} To analyze the effect of capital control measures on global fund inflows, we use the CCM dataset by focusing only on interventions classified with an “Inward” action direction. We further examine the impact of different levels of action intensity across various policy categories. These identified events serve as the basis for implementing an event study analysis on fund flow dynamics.

\noindent\paragraph{Control Variables} Building on capital flow theory \citep{broner2013gross, erten2021capital,tille2010international}, we incorporate both global push factors and domestic pull factors as key determinants of cross-border capital movements. Global push factors capture external conditions that influence investors' decisions across countries, including global risk sentiment (measured by the VIX), U.S. interest rates, and other indicators of global financial conditions. In contrast, domestic pull factors reflect country-specific fundamentals that attract capital, such as economic growth, financial openness, exchange rate regime, and the presence or stringency of capital control policies. These drivers have been widely emphasized in the literature on capital flows, including foundational work by \citet{RePEc:pra:mprapa:7125} and more recent contributions by \citet{RePEc:eee:jimfin:v:48:y:2014:i:pb:p:221-248}.

\subsection{Empirical Method and Results}

\noindent\paragraph{Event Study} As a preliminary application in this paper, we use China, Australia and the US to conduct an event study analysis examining whether inward capital control measures have impact on fund inflows into the country. To evaluate the dynamic effects of capital control measures on cross-border fund flows, we implement an event study framework using monthly fund-level data. We focus on policy interventions classified as \textit{inward} capital controls and further distinguish events by their \textit{action intensity}---namely, \textit{liberalizing} versus \textit{restrictive and conditional} actions.

For each capital control event, we define an event window of 13 months, spanning from six months before to six months after the policy implementation date. The dependent variable is the percentage change in fund flow relative to assets at the beginning of the month, denoted as \texttt{flowpct}. We assign an event time variable $t \in [-6, +6]$ for each observation in the window, where $t = 0$ indicates the month of the policy event.

The main regression specification is as follows:

\begin{equation}
\texttt{Flowpct}_{i t} = \sum_{\tau = -6}^{+6} \sum_{k \in \{\text{L}, \text{R}\}} \beta_{\tau}^k \cdot \mathbb{1}\{t = \tau \text{ and } \texttt{Intensity} = k\} + \alpha_i + \gamma_t + \varepsilon_{i t}
\label{eq:eventstudy}
\end{equation}

where $\mathbb{1}\{\cdot\}$ is an indicator function for event time $\tau$ and action intensity $k$. $\alpha_i$ denotes fund fixed effects, and $\gamma_t$ captures global month fixed effects. The coefficients $\beta_{\tau}^k$ trace the average treatment effect of each policy type over time. Standard errors are clustered at the fund or country level. For descriptive analysis, we additionally plot the average \texttt{flowpct} over the event window by policy type, with 95\% confidence intervals constructed using standard errors of the mean. The event study plot shows in follows.

\begin{figure}[H]
    \centering
    \includegraphics[width=0.9\textwidth]{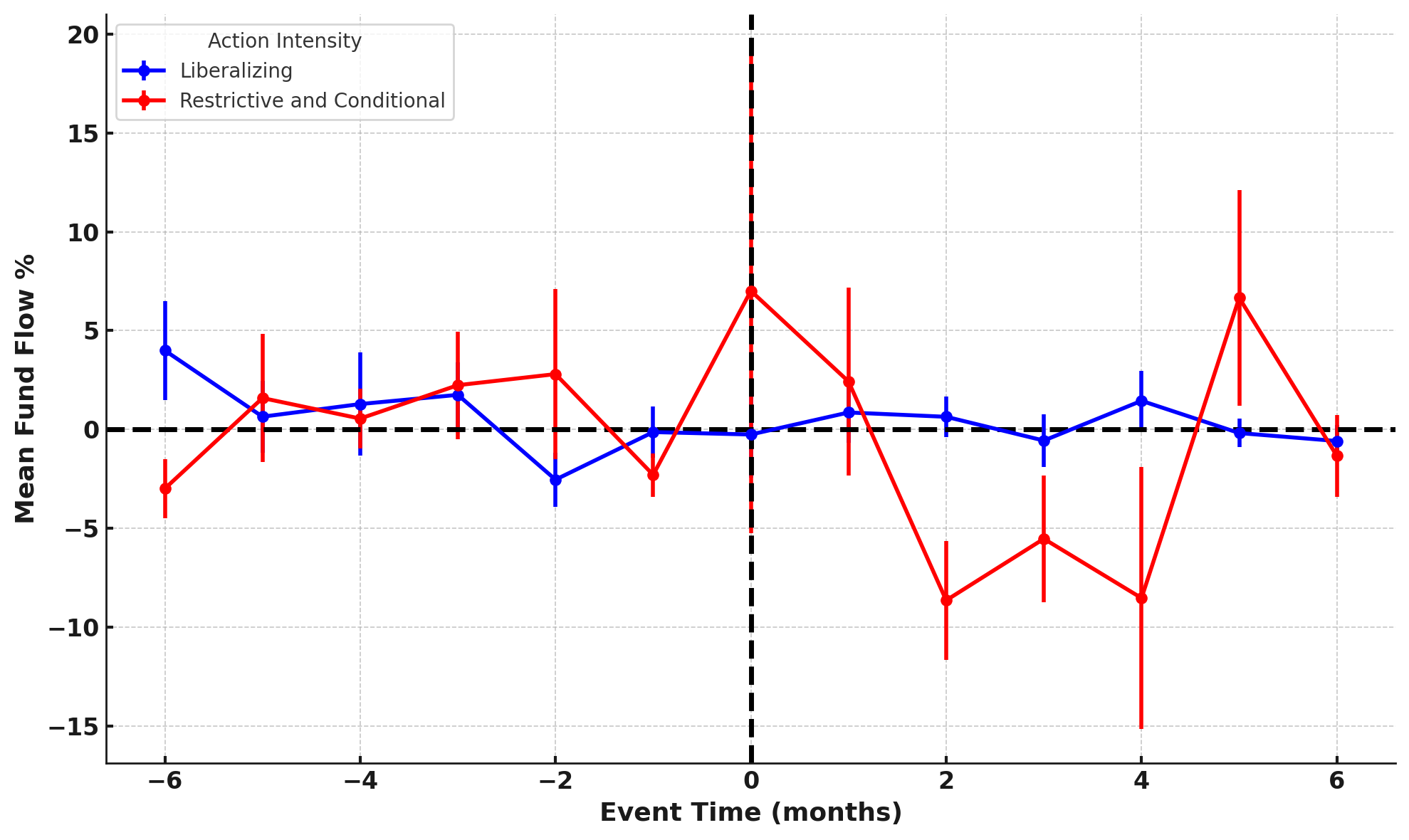}
    \caption{Event Study Analysis: A Case Study of China}
    \label{fig:eventstudy}
\end{figure}

For the case study of China in Figure~\ref{fig:eventstudy}, the red line represents the average fund flow percentage around the implementation of restrictive or conditional capital inflow measures, while the blue line corresponds to liberalizing interventions. The event study reveals notable patterns. Restrictive and conditional inward capital control policies introduced by China are associated with a significant decline in fund inflows: Before implementing capital control policies, the average fund flow into China was quite stable. Starting approximately one month after implementation, the average fund flow percentage drops, reaching a cumulative decline of around 8\% within three months. In contrast, liberalizing policies exhibit only a modest effect on capital inflows after implementing the loosening policies, and last for 2 months. 

\begin{figure}[H]
    \centering
    \includegraphics[width=0.9\textwidth]{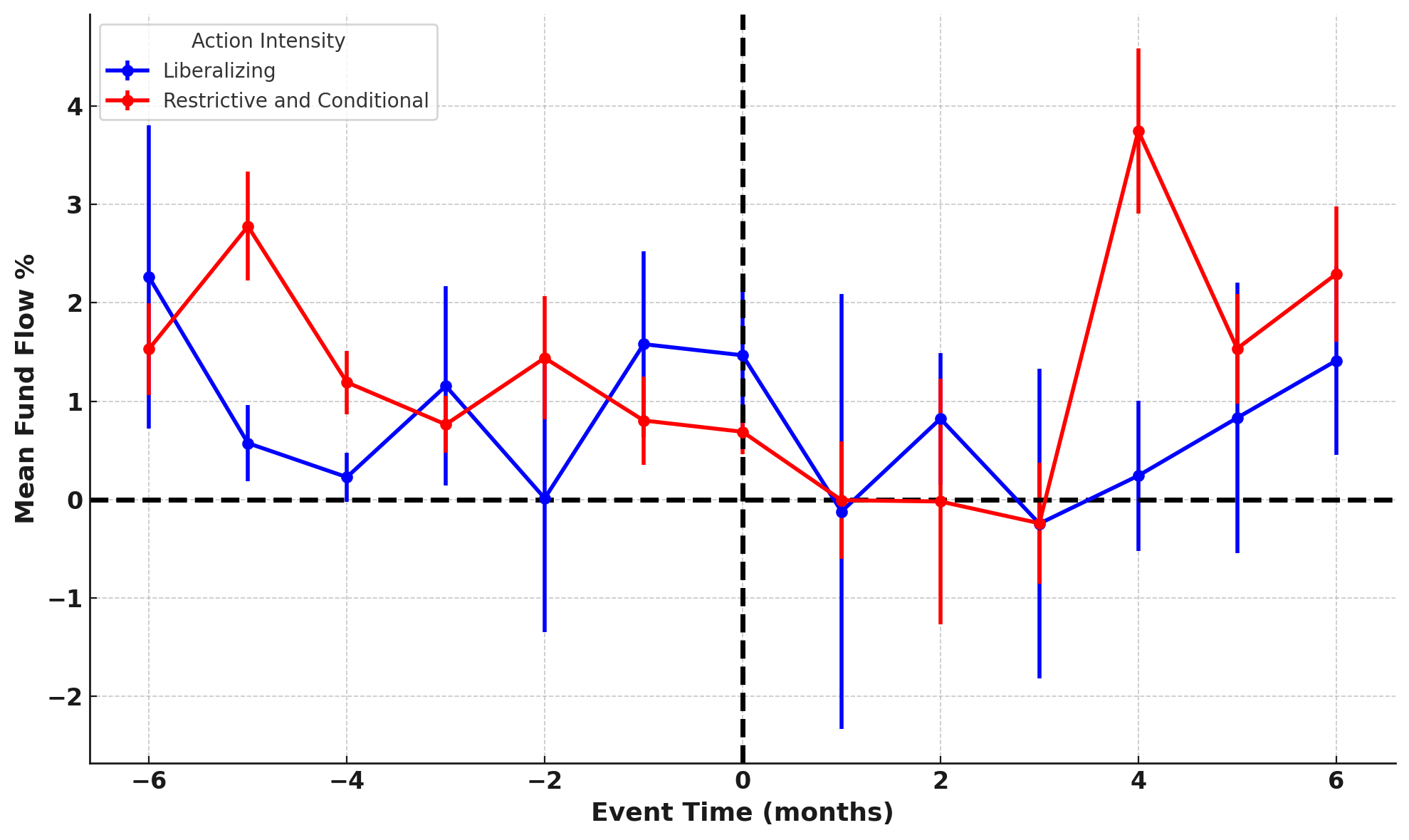}
    \caption{Event Study Analysis: A Case Study of Australia}
    \label{fig:eventstudy_aus}
\end{figure}

\begin{figure}[H]
    \centering
    \includegraphics[width=0.9\textwidth]{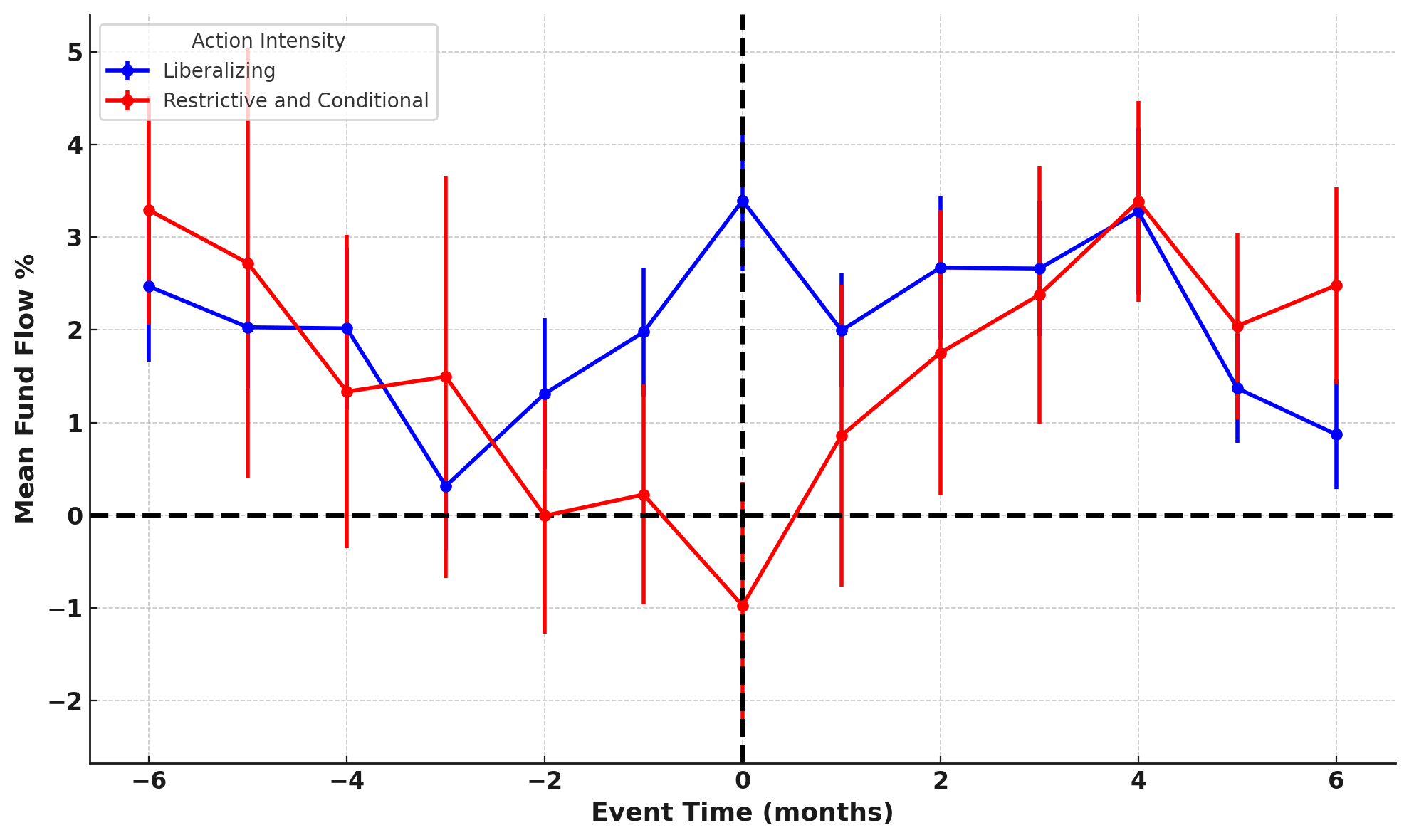}
    \caption{Event Study Analysis: A Case Study of the US}
    \label{fig:eventstudy_us}
\end{figure}

The case of Australia and the US exhibits a distinct pattern, as shown in Figure~\ref{fig:eventstudy_aus} and Figure \ref{fig:eventstudy_us}. Fund flows were significantly positive in the months preceding the implementation of restrictive and conditional capital control measures. However, the effects taper off and become statistically insignificant immediately after the policy interventions, remaining close to zero for several months. Notably, a significant reflow of capital into Australia is observed approximately four months after the intervention. This suggests that the impact of restrictive capital controls in Australia persists for up to four months. In contrast, liberalizing capital account interventions appear to have no significant effect on boosting fund inflows upon implementation.

A particularly interesting pattern regarding action intensity emerges in the case of the United States. The effects of liberalizing and restrictive measures exhibit opposite dynamics around the time of implementation. Inward-oriented restrictive capital control measures show a statistically significant impact only in the month of implementation, after which capital inflows remain positive and continue on an upward trend. Surprisingly, liberalizing measures also produce a short-term significant increase in fund inflows; however, this effect is temporary, as inflows revert to their pre-intervention levels shortly thereafter.\footnote{This figure presents a preliminary example of the event study results. To assess robustness, additional methods and specifications can be employed to validate the effects of capital control measures. Due to space constraints, we leave a detailed robustness analysis to a separate paper.} These event-level findings underscore the potential for future research to further investigate the nuanced and heterogeneous effects of capital control intervention.

\section{Conclusion\label{sec:conclusion}}

This paper develops a new event-level Capital Control Measures (CCM) dataset based on the IMF’s AREAER reports, leveraging prompt-based large language models (GPT-4.1 \cite{gpt412025}) and modern natural language processing (NLP) techniques. Using textual analysis, we enrich each capital control intervention with multidimensional characteristics, including action type, action intensity, action direction, implementing actor, target country, target industry, and several binary indicators such as whether the measure is related to trade policy, national security, or sanctions. The stylized facts derived from the CCM dataset reveal distinct patterns around major global episodes, such as the 2008 financial crisis and the COVID-19 pandemic, and highlight heterogeneity across different income and regional country groups. Applying the CCM dataset in an empirical setting, we further explore how capital control interventions affect global fund flows using an event study framework. Our preliminary findings from China, Australia, and the United States suggest that restrictive capital control measures tend to have a more significant impact on fund inflows compared to liberalizing measures, which exhibit only mild effects in attracting capital. Moreover, we observe notable heterogeneity across countries, indicating that the effectiveness of capital control interventions may vary depending on national context, institutional frameworks, and market conditions, which is consistent with existing literature.

To automate the categorization process and develop a model that generalizes to other sources of capital control events, we finetune an open-sourced large language model based on Meta Llama 3.1 \citep{grattafiori2024llama}. We refer to the resulting model as \textit{CCM-Llama}. To support this effort, we construct a rigorous data collection pipeline and compile a large-scale finetuning dataset comprising 29,012 examples. Our finetuned model achieves 99.55\% accuracy on the binary classification task and up to 90.09\% accuracy on the hierarchical classification task, outperforming all baselines including the state-of-the-art commercial model. A key advantage of \textit{CCM-Llama} lies in its ability to incorporate multiple sources of policy measures and to provide real-time classification of capital control events from external sources such as market news and central bank announcements. This capability enables researchers and investors to continuously monitor and analyze the evolving dynamics of global financial markets.

%%%%%%%%%%%%%%%%%%%%%%%%%%%%%%%%%%%%%%%%%%%%%%%%%
\clearpage
\begin{singlespace}
\bibliographystyle{aer}
\bibliography{our-cites.bib}
\end{singlespace}
%%%%%%%%%%%%%%%%%%%%%%%%%%%%%%%%%%%%%%%%%%%%%%%%%

%%%%%%%%%%%%%%%%%%%%%%%%%%%%%%%%%%%%%%%%%%%%%%%%%
%%%%% These commands start the appendix and change the Table & Figure numbering
\newpage
\appendix
\setcounter{table}{0}
\renewcommand{\tablename}{Appendix Table}
\renewcommand{\figurename}{Appendix Figure}
\renewcommand{\thetable}{A\arabic{table}}
\setcounter{figure}{0}
\renewcommand{\thefigure}{A\arabic{figure}}
%%%%%%%%%%%%%%%%%%%%%%%%%%%%%%%%%%%%%%%%%%%%%%%%%

% \section{Appendix Tables and Figures}
% % \input{tab_tex/other-regressions.tex}

% \newpage 
% \section{Appendix One \label{sec:appendix:first}}
% \renewcommand{\thetable}{B\arabic{table}}
% \setcounter{table}{0}
% \renewcommand{\thefigure}{B\arabic{figure}}
% \setcounter{figure}{0}

\appendix
\section{Description of Capital Control Categories}
\label{app:cc_des}
\definecolor{lightgray}{gray}{0.95}

\begin{longtable}{@{}p{2.5cm}p{5.5cm}p{6cm}@{}}
\caption{Capital Control Category Descriptions (Multi-Page)}%
\label{tab:cc_des} \\
\toprule
\textbf{Index} & \textbf{Category} & \textbf{Description} \\
\midrule
\endfirsthead

\toprule
\textbf{Index} & \textbf{Category} & \textbf{Description} \\
\midrule
\endhead

% Footer for all but the last page
\midrule
\multicolumn{3}{r}{\emph{Continued on next page}} \\
\bottomrule
\endfoot

% Footer for the last page
\bottomrule
\endlastfoot

% ACTIVATE ROW COLORING NOW
\rowcolors{1}{white}{lightgray}

% Paste all your rows below this line. Example:
XI.A. & Controls on capital transactions [ka] & Restrictions on cross-border capital flows, covering investment, credit, and real estate transactions. \\
XI.A.2. & Controls on capital and money market instruments & Rules regulating international transactions in equity, debt, money markets, and investment funds. \\
XI.A.2.a. & On capital market securities & Measures on shares and bonds issued or acquired by residents or nonresidents. \\
XI.A.2.a.1. & Shares or other securities of a participating nature [eq] & Controls on equity investments involving ownership or participation rights. \\
XI.A.2.a.1.i & Purchase locally by nonresidents [eq\_plbn] & Restrictions on foreigners buying domestic equity locally. \\
XI.A.2.a.1.iv & Sale or issue abroad by residents [eq\_siar] & Controls on residents issuing or selling equity abroad. \\
XI.A.2.a.1.iii & Purchase abroad by residents [eq\_pabr] & Measures on residents buying equity in foreign markets. \\
XI.A.2.a.1.ii & Sale or issue locally by nonresidents [eq\_siln] & Regulations on foreigners issuing equity locally. \\
XI.A.2.a.2. & Bonds or other debt securities [bo] & Controls on transactions in debt instruments like bonds and notes. \\
XI.A.2.a.2.i & Purchase locally by nonresidents [bo\_plbn] & Restrictions on foreigners purchasing domestic bonds. \\
XI.A.2.a.2.iv & Sale or issue abroad by residents [bo\_siar] & Controls on residents issuing debt securities abroad. \\
XI.A.2.a.2.iii & Purchase abroad by residents [bo\_pabr] & Measures on residents buying foreign bonds. \\
XI.A.2.a.2.ii & Sale or issue locally by nonresidents [bo\_siln] & Regulations on foreign entities issuing debt locally. \\
XI.A.2.b. & On money market instruments [mm] & Rules concerning short-term securities such as T-bills and commercial paper. \\
XI.A.2.b.1 & Purchase locally by nonresidents [mm\_plbn] & Controls on foreigners buying domestic money market instruments. \\
XI.A.2.b.4 & Sale or issue abroad by residents [mm\_siar] & Restrictions on residents issuing money market instruments abroad. \\
XI.A.2.b.3 & Purchase abroad by residents [mm\_pabr] & Regulations on residents acquiring money market instruments abroad. \\
XI.A.2.b.2 & Sale or issue locally by nonresidents [mm\_siln] & Measures on nonresidents issuing short-term instruments domestically. \\
XI.A.2.c. & On collective investment securities [ci] & Controls on cross-border transactions in mutual funds and similar vehicles. \\
XI.A.2.c.3 & By residents to nonresidents [cio] & Restrictions on domestic collective vehicles selling to nonresidents. \\
XI.A.2.c.1 & By nonresidents to residents [cii] & Controls on foreign investment vehicles marketed to residents. \\
XI.A.3. & Controls on derivatives and other instruments & Measures regulating cross-border transactions in financial derivatives. \\
XI.A.3.a. & Purchase locally by nonresidents & Restrictions on nonresidents buying derivatives locally. \\
XI.A.3.b. & Sale or issue locally by nonresidents & Regulations on foreigners issuing derivatives domestically. \\
XI.A.3.c. & Purchase abroad by residents & Controls on residents acquiring foreign derivatives. \\
XI.A.3.d. & Sale or issue abroad by residents & Restrictions on residents issuing derivatives overseas. \\
XI.A.4. & Controls on credit operations & Rules on lending, guarantees, and financial credit between residents and nonresidents. \\
XI.A.4.b. & Financial credits [fc] & Regulations on cross-border financial loans and credit lines. \\
XI.A.4.b.1 & By residents to nonresidents [fco] & Controls on residents providing financial loans abroad. \\
XI.A.4.b.2 & By nonresidents to residents [fci] & Measures on foreign entities lending to domestic parties. \\
XI.A.4.a. & Commercial credits & Measures on trade-related deferred payment and credit agreements. \\
XI.A.4.a.1 & By residents to nonresidents & Restrictions on trade credits extended by residents abroad. \\
XI.A.4.a.2 & To residents from nonresidents & Controls on commercial credit provided by foreign parties. \\
XI.A.4.c. & Guarantees, sureties, and financial backup facilities & Controls on financial guarantees supporting cross-border financial operations. \\
XI.A.4.c.1 & By residents to nonresidents & Restrictions on guarantees extended by residents to foreigners. \\
XI.A.4.c.2 & To residents from nonresidents & Measures on financial backing offered to residents by nonresidents. \\
XI.A.7. & Controls on real estate transactions & Measures restricting cross-border property ownership or sales. \\
XI.A.7.a. & Purchase abroad by residents & Restrictions on residents buying property overseas. \\
XI.A.7.b. & Purchase locally by nonresidents & Controls on foreigners acquiring domestic real estate. \\
XI.A.7.c. & Sale locally by nonresidents & Regulations on foreign owners selling property domestically. \\
XI.A.5. & Controls on direct investment [di] & Rules managing long-term cross-border investments involving control or influence. \\
XI.A.5.a & Outward investment [dio] & Controls on domestic entities investing directly abroad. \\
XI.A.5.b & Inward direct investment [dii] & Measures on foreign direct investment into domestic firms. \\
XI.A.5.c & Liquidation of direct investment [ldi] & Regulations on the repatriation of capital from divested investments. \\
\end{longtable}

\section{AREAER Country Mapping}
\label{app:country_mapping}
\begin{table}[!t]
    \centering
    \caption{Country Name Mappings}
    \begin{tabular}{ll}
        \toprule
        \textbf{Original Name} & \textbf{Standardized Name} \\
        \midrule
        Türkiye & Turkey \\
        Hong Kong SAR & Hong Kong \\
        Hong Kong Special Administrative Region & Hong Kong \\
        Democratic Republic of the Congo (DRC) & Democratic Republic of the Congo \\
        Democratic Republic of the Congo & Democratic Republic of the Congo \\
        Republic of Korea & South Korea \\
        Korea & South Korea \\
        Islamic Republic of Iran & Iran \\
        Republic of Yemen & Yemen \\
        Syrian Arab Republic & Syria \\
        Lao P.D.R. & Laos \\
        Lao People’s Democratic Republic & Laos \\
        People’s Republic of China & China \\
        People’s Republic of China—Hong Kong SAR & Hong Kong \\
        Swaziland & Eswatini \\
        Kingdom of Eswatini & Eswatini \\
        Cape Verde & Cabo Verde \\
        República Bolivariana de Venezuela & Venezuela \\
        República Bolivariana De Venezuela & Venezuela \\
        Russian Federation & Russia \\
        Papua New-Guinea & Papua New Guinea \\
        Federated States of Micronesia & Micronesia \\
        Serbia and Montenegro & Serbia \\
        Republic of Serbia & Serbia \\
        Republic of Montenegro & Montenegro \\
        Islamic Republic of Afghanistan & Afghanistan \\
        Islamic State of Afghanistan & Afghanistan \\
        Czech Republic & Czechia \\
        Former Yugoslav Republic of Macedonia & North Macedonia \\
        Republic of North Macedonia & North Macedonia \\
        Republic of Congo (Congo) & Republic of Congo \\
        Democratic Republic of Timor-Leste & Timor-Leste \\
        Republic of Azerbaijan & Azerbaijan \\
        Republic of Fiji & Fiji \\
        Socialist People’s Libyan Arab Jamahiriya & Libya \\
        \bottomrule
    \end{tabular}
    \label{tab:country_mapping}
\end{table}

\end{document}